\documentclass{aa}
\usepackage{natbib}
\bibpunct{(}{)}{;}{a}{}{,} 
\usepackage{graphicx}
\usepackage{txfonts}

\newcommand{\be}{\begin{equation}}
\newcommand{\ee}{\end{equation}}
\newcommand{\bea}{\begin{eqnarray}}
\newcommand{\eea}{\end{eqnarray}}
\newcommand{\bann}{\begin{eqnarray*}}
\newcommand{\eann}{\end{eqnarray*}}
\newcommand{\bi}{\begin{itemize}}
\newcommand{\ei}{\end{itemize}}

\newcommand{\bcen}{\begin{center}}
\newcommand{\ecen}{\end{center}}
\newcommand{\rxdieciocho}{\hbox{RX~J1856.5$-$3754~}}
\newcommand{\rxcerosiete}{\hbox{RX~J0720.4$-$3125~}}
\newcommand{\rbdoce}{\hbox{RBS~1223~}}
\begin{document}
\title{2D Cooling of Magnetized Neutron Stars}

   \author{Deborah~N.~Aguilera\inst{1,2}
\and Jos\'e A. Pons \inst{1} 
\and Juan A. Miralles \inst{1}
          }
\offprints{aguilera@tandar.cnea.gov.ar, jose.pons@ua.es}  
  \institute{Department of Applied Physics, University of Alicante,
             Apartado de Correos 99, E-03080 Alicante, Spain\\
\and
Theoretical Physics, Tandar Laboratory, 
 National Council on Atomic Energy (CNEA-CONICET), Av. Gral Paz 1499, 1650 San Mart\'{\i}n, 
 Buenos Aires, Argentina}
    \date{Received: date / Revised version: date}
\abstract
   {Many thermally emitting, isolated 
neutron stars have magnetic fields that are larger than $10^{13}$~G. A 
realistic cooling model that includes the presence of high magnetic fields should be reconsidered.}
{We investigate 
the effects of an anisotropic
temperature distribution and Joule heating on the cooling 
of magnetized neutron stars.}
   {The 2D heat transfer equation with anisotropic thermal 
conductivity tensor and including all relevant
   neutrino emission processes is solved for realistic models of the neutron star interior and crust.}
   {The presence of the magnetic field affects significantly the thermal surface distribution and the
   cooling history during both, the early neutrino cooling era and the late photon cooling era.}
   { There is a large effect of 
 Joule heating on the thermal evolution of strongly magnetized neutron stars.
Both magnetic fields and Joule heating play an important role in keeping 
magnetars warm for a long time. 
 Moreover, this effect is important for
intermediate field neutron stars 
and should be considered in radio--quiet isolated neutron stars or high magnetic field
radio--pulsars.}
   \maketitle
   
\section{Introduction}

Observation of thermal emission from neutron stars (NSs) 
can provide not only information on 
the physical properties such as the magnetic field, temperature, and chemical composition of 
the regions where this radiation is produced but also information on the 
properties of matter at higher densities deeper inside the star \citep{Yakovlev2004,Page2006}. 
To derive this
information, we need to calculate the structure and evolution of the star, 
and compare the theoretical model with the observational data.
Most previous studies assumed a spherically symmetric 
temperature distribution. 
However, there is increasing evidence that this is not the case for 
most nearby neutron stars whose thermal emission is visible in the
X-ray band of the electromagnetic spectrum  \citep{Zavlin2007,Haberl2007}.
The anisotropic temperature distribution 
 may be produced not only in the low density regions where the spectrum
is formed and preliminary investigations had focused their attention,   
but also in intermediate density regions, such as the solid crust, where a complicated 
magnetic field geometry could cause a coupled magneto-thermal 
evolution. In some extreme cases, this anisotropy may even be present in the
poorly known interior, where neutrino processes are responsible  for the energy removal. 

The observational fact that most thermally emitting isolated 
NSs have magnetic fields larger than $10^{13}$ G \citep{Haberl2007}, which 
is sometimes confirmed by spin down measurements,
leads to the conclusion that a realistic cooling model must
not avoid the inclusion of the effects produced by the presence of 
high magnetic fields. The transport processes that occur in the 
interior are affected by these strong magnetic fields and their effects 
are expected to have observable consequences, 
in particular for highly magnetized NSs or magnetars. Moreover,  
the large surface magnetic field strengths inferred from the observations 
probably indicate that the interior field could reach even larger values,
as theoretically predicted by some models \citep{TD1993}. 

The presence of a magnetic field affects the transport properties 
of all plasma components, especially the electrons.
In general, the motion of free electrons perpendicular to the 
magnetic field is quantized in Landau levels, and
the thermal and electrical conductivities exhibit quantum oscillations.
In the limit of a strongly quantizing field, in which almost all electrons populate 
the lowest level, such as in the envelope of a NS, a quantum description is
necessary to calculate the thermal and electrical conductivities. 
Earlier calculations by 
\cite{Canuto1970} and \cite{Itoh1975} concluded 
that the electron thermal conductivity is strongly suppressed in the 
direction perpendicular to the magnetic field and increased
along the magnetic field lines, which reduces the thermal insulation of the envelope 
({\it heat blanketing}). 
Thus, there is an anisotropic heat transport in the NS's envelope 
governed by the magnetic field geometry,  that produces
a non-uniform surface temperature. 

The anisotropy in the surface temperature of a NS
appears to be confirmed by the analysis of observational data 
from isolated NSs (see \cite{Zavlin2007} and \cite{Haberl2007} 
for reviews on the current status 
of theory and observations). The mismatch between the extrapolation 
to low energy of fits to the X-ray spectra,
and the observed Rayleigh-Jeans tail in the optical band ({\it optical excess flux}),
cannot be addressed using a unique temperature. 
Several simultaneous fits to multiwavelength spectra of
 \rxdieciocho \citep{Pons2002,Truemper2004}, 
\rbdoce \citep{Schwope2005,Schwope2007},  and 
\rxcerosiete \citep{Perez2006} are explained by a small hot emitting 
area $\simeq$ 10--20 km$^2$, and an extended cooler component.    
Another piece of evidence that strongly supports the nonuniform temperature distribution are  
pulsations in the X-ray signal  of some objects of amplitudes $\simeq$ 5--30 $\%$, 
some of which have 
irregular light curves that point towards a non-dipolar temperature distribution.
All of these facts reveal that the idealized picture of a NS with a dipolar magnetic 
field  and uniform surface temperature is oversimplified. 

In a pioneering work, \cite{Greenstein1983} obtained
the temperature at the surface of a NS as a function 
of the magnetic field inclination angle in a simplified plane-parallel approximation. 
This model was applied to different magnetic field configurations and the observational 
consequences of a non-uniform temperature 
distribution were analyzed in the pulsars Vela and Geminga among 
others \citep{Page1995}.
\cite{Potekhin2001} 
improved the former calculations including
realistic thermal conductivities. 
Nevertheless, the temperature anisotropy as generated in the envelope 
may be insufficiently to be consistent with the observed 
thermal distribution and, in this case, should originate 
deeper inside the NS \citep{Geppert2004,Azorin2006}. 

Crustal confined magnetic fields could be responsible for the 
surface thermal anisotropy. 
In the crust,  even if a strong magnetic field is present, 
the electrons occupy a large number of Landau levels 
and the classical approximation remains valid during a long 
time in the thermal evolution. The magnetic field limits the movement of electrons
in the direction perpendicular to the field and, 
since they are the main carriers
of the heat transport, the thermal conductivity 
in this direction is highly suppressed, while remaining almost unaffected 
along the field lines. 
Temperature distributions in the crust were obtained as 
stationary solutions of the diffusion equation with axial symmetry \citep{Geppert2004}. 
The approach assumes an isothermal core and a magnetized envelope as an inner and 
outer boundary condition, respectively. 
The results show important deviations from the crust isothermal case for crustal confined magnetic 
fields with strengths larger than $10^{13}$ G and temperatures below $10^{8}$ K. 
Similar conclusions were obtained
considering not only poloidal but also toroidal components for the magnetic field
\citep{Azorin2006, Geppert2006}. 
This models succeeded in explaining simultaneously the observed X-ray spectrum, the optical excess,
the pulsed fraction, and other spectral features for some isolated NS 
such as \rxcerosiete \citep{Perez2006} and \rxdieciocho  \citep{Geppert2006}. 

Non-uniform surface temperature in NSs
was studied by different authors using simplified models \citep{ShibYak1996,Potekhin2001}. 
Although these models can provide useful information, a detailed 
investigation of heat transport in 2D must be completed to obtain more solid conclusions.
However, this is not the only effect that must be revisited to study the cooling of NSs.
For isolated NSs, different relevant magnetic field dissipation processes were 
identified \citep{Goldreich1992}. The {\it Ohmic} dissipation rate is determined by 
the finite conductivity of the constituent matter. In the crust, the electrical resistivity 
is mainly due to electron-phonon and electron-impurity scattering processes
\citep{Flowers1976}, resulting in more efficient Ohmic dissipation than in the fluid interior. 
The strong temperature dependence of the resistivity leads to rapid dissipation of the 
magnetic energy in the outermost low-density regions during the early evolution of a hot NS,
which becomes less relevant as the star cools down. 
Joule heating in the crustal layers due to Ohmic decay was thought to affect only
the late photon cooling era in old NS ($\geq 10^7$ yr), and to be an efficient mechanism 
to maintain the surface temperature as high as $\simeq 10^{4-5}$ K for a long time
\citep{Miralles98}. \cite{Page2000} studied the 1-D thermal evolution of NSs
combined with an evolving Stokes function that defines a purely poloidal, 
dipolar magnetic field. 
The Joule heating rate was evaluated averaging the currents over the azimuthal angle. 
However, for strongly magnetized NS, Joule heating 
can be important much earlier in the evolution.
In a recent work, \cite{Kaminker2006} placed a heat source inside the outer crust 
of a young, warm, magnetar of field strength $5\times10^{14}$ G. 
To explain observations, they 
concluded that the heat source should be located at a density 
$\lesssim 5\times 10^{11}$ g~cm$^{-3}$, 
and the heating rate should be $\gtrsim 10^{20}$ erg~cm$^{-3}$~s$^{-1}$ for at least 
$5\times10^4$ yr. 
Anisotropic heat transport 
is neglected in these simulations, which were performed in spherical symmetry, 
assuming that 
it will  not affect the results in the early evolution. 
Nevertheless we will show that, in 2D simulations, the effect of anisotropic
heat transport is important.

In addition to purely Ohmic dissipation, strongly magnetized NSs can also experience a
{\it Hall drift} with a drift velocity proportional to the magnetic field strength. 
Although the Hall drift conserves the magnetic energy 
and it is not a dissipative mechanism by itself, it can enhance the Ohmic decay by  compressing
the field into small scales, or by displacing currents to regions with higher resistivity,
where Ohmic dissipation is more efficient. 
Recently, the first 2D-long term simulations of the magnetic field evolution
in the crust studied the interplay of Ohmic dissipation and the Hall drift effect 
\citep{PonsGeppert2007}. 
It was shown that, for magnetar field strength, the characteristic timescale during which 
Hall drift influences Ohmic dissipation is of about $10^{4}$ yr. All of these studies 
imply
that both field decay and Joule heating play a role in 
the cooling of neutron stars born with field strengths $\geq 10^{13}$ G.

We will show that, during the neutrino cooling era and the early stages of the photon cooling era, 
the thermal evolution is coupled to the magnetic field decay, since both cooling and magnetic
field diffusion proceed on a similar timescale ($\approx 10^{6} $ yr). 
The energy released by magnetic field decay in the crust
could be an important heat source that modifies or even controls the thermal evolution of a NS.
Observational evidence of this fact  is shown in \cite{PonsLink2007}. They found a 
strong correlation between the inferred 
magnetic field and the surface temperature for a wide range of magnetic fields: 
from magnetars ($\geq 10^{14}$ G), through radio-quiet isolated neutron stars 
($\simeq 10^{13}$ G) 
down to some ordinary pulsars ($\leq 10^{13}$ G). The main conclusion is that,
rather independently from the stellar structure and the matter composition,
the correlation can be explained by the decay of currents
on a timescale of $\simeq 10^{6}$ yr.

The aim of the present work is to study in a more consistent way the cooling of a 
realistic NS under the effects of large magnetic fields, including the effects of an anisotropic
temperature distribution and Joule heating in 2D simulations.
As a first step towards a fully coupled magneto-thermal evolution, a phenomenological 
law for the magnetic field decay is considered. 

This article is structured as follows.
In Sect.~2 we discuss
the equations governing the magnetic field structure and evolution, while Sect.~3 is devoted to
the thermal evolution equations. Sect.~4 presents the microphysics inputs. 
Sect.~5 and 6
contain our results for weakly and strongly magnetized NSs, respectively. In Sect. 7, we focus 
on the effects of field decay and Joule heating on the cooling history of a NS. 
Finally, in Sect. 8 we present the 
main conclusions and perspectives of the present work. 

\section{Magnetic field structure and evolution.} 
\label{SecMagnetic}
 
While the large-scale external structure of the magnetic field of NSs is usually  
represented by the vacuum solution of an external dipole, or sometimes a more complex 
magnetosphere, the structure of the magnetic field in the interior of  NSs is poorly known. 
Results from MHD  simulations show that stable configurations require 
the coexistence of both poloidal and toroidal components, approximately of the same strength  
\citep{Braithwaite2004}, although predominantly poloidal configurations may be stabilized 
by rapid rotation \citep{Rheinhardt2006}. In general, a realistic NS magnetic field model should  
contain both components, and their location and relative strength should vary.
Moreover, two-dimensional simulations \citep{PonsGeppert2007} showed that, 
while the initial magnetic field 
configuration determines the early evolution of the field ($t<10^4$ yr), 
at later stages a more stable configuration, consisting of a dipolar poloidal component
and a higher order toroidal component, is preferred. 

We consider the Newtonian approximation of a NS magnetic field because 
general relativistic corrections are not important in our study. 
In axial symmetry, the magnetic field 
can be decomposed into poloidal and toroidal components \citep{Raedler2000}  
\bea
\vec{B}=\vec{B}_{\rm pol} + \vec{B}_{\rm tor}~, 
\eea
which are represented, respectively, by two scalar functions $\tilde{\mathcal S}$, 
$\tilde{\mathcal T}$: 
\bea 
\vec{B}_{\rm pol} &=& \nabla \times (\vec r \times \nabla \tilde{\mathcal S})~\\ 
\vec{B}_{\rm tor} &=& - \vec{r} \times \nabla \tilde{\mathcal T}   
\eea 
Here, $\tilde{\mathcal S}$ and $\tilde{\mathcal T}$ depend on the 
spherical coordinates  $r,\theta$, and $\vec r$ is a radial vector.  
 
Expanding the angular part of the scalar functions in Legendre polynomials, 
we can write 
\bea 
\tilde{\mathcal S}(r,\theta)= C \sum_l \frac{P_l(\cos \theta)}{r} {\mathcal S}_l(r,t)~, \nonumber \\ 
\tilde{\mathcal T}(r,\theta)= C \sum_l \frac{P_l(\cos \theta)}{r} {\mathcal T}_l(r,t)~, 
\label{STdef} 
\eea 
where $P_l(\cos \theta)$ is the Legendre polynomial of order $l$ and $C$
is a normalization constant. 
For $l=1$,  wich represents dipolar fields, after 
normalizing the field to its surface value at the magnetic pole, $B$,
($C={R_{NS}^2 B/2}$) 
and the radial coordinate to the NS radius ($x=r/R_{NS})$,
the components of the magnetic field can be written in terms of  
the two functions, ${\mathcal S}_1 (x,t)$ and ${\mathcal T}_1 (x,t)$, as follows 
\bea 
B_{r} &=& B \frac{\cos{\theta}}{x^{2}}  {\mathcal S}_1 (x,t) \nonumber \\  
B_{\theta} &=& - B \frac{\sin {\theta}}{2 x} \frac{\partial {\mathcal S}_1 (x,t)}{\partial x} \nonumber\\ 
B_{\phi}&=& B \frac{\sin {\theta}}{2 x} {\mathcal T}_1 (x,t)~, 
\label{bform} 
\eea 
where in the following we omit the subindex ($l=1$) for clarity.
We note that ${\mathcal S}(x,t)$ is normalized such that it reaches the value  
of 1 at the surface. 
These two arbitrary functions are subject to suitable boundary conditions. 
To match continuously the external vacuum dipole solution, for example,  
${\mathcal S}(x,t)$ must satisfy 
$\partial {\mathcal S}(x,t)/\partial x =-{\mathcal S}(x,t)$, at $x=1$. 
Other boundary conditions are discussed below. 
 
\subsection{Magnetic field geometry} 
 
From the above general form of the magnetic field components, different interesting  
cases can easily be recovered. We describe three possible configurations 
that we explored in this work. 
 
\subsubsection{Force-free fields (FF model)} 
\label{secff}
 
One of the particular models we consider here are the force-free fields. 
They satisfy: 
\bea 
\vec{\nabla} \times \vec{B} &=& \mu \vec{B}, \quad\quad 
\vec{B} \cdot \vec{\nabla} \mu = 0 \quad, 
\label{maxwell3} 
\eea 
where $\mu$ is a parameter related to the magnetic field curvature, which naively
 can be interpreted as a wavenumber of the Stokes function $\mathcal S$.  
For simplicity, we consider solutions with $\mu=$constant such that the second equation  
is satisfied automatically. 
A general interior solution that  fulfils the equality between the two components ($r, \theta$) 
in the first equation can be obtained by choosing  
\bea 
{\mathcal T}(x,t) = \mu {\mathcal S}(x,t),  
\label{vectorAff} 
\eea 
Factoring the time dependence in an arbitrary function, ${\mathcal S}(x,t)=f(t) A(x)$, 
the $\phi$--component of the first equality in Eq. \ref{maxwell3} produces  a   
form of the Riccati-Bessel equation for $A(x)$ whose solution can be written analytically 
in terms of the spherical Bessel functions of the first and second kind. For $l=1$, we have 
\bea 
A(x) &=& a \hat{x} j_1( \hat{x}) + b \hat{x} n_1(\hat{x})~, \nonumber\\ 
j_1(\hat x) &=& \frac{\sin{\hat x}}{\hat x^2} - \frac{\cos{\hat x}}{\hat x}~, \nonumber \\ 
n_1(\hat x) &=& -\frac{\cos{\hat x}}{\hat x^2} - \frac{\sin{\hat x}}{\hat x} 
\label{Afunc}
\eea 
where $\hat x=\mu R_{\rm NS} x$.   
From this, the magnetic field is given by
\bea 
B_r &=& B \frac{\cos \theta}{x^2} A(x), \nonumber \\  
B_{\theta} &=& - B \frac{\sin \theta}{2 x}\frac{d A(x)}{d x}, \nonumber \\ 
B_\phi &=& B \mu R_{\rm NS} \frac{\sin \theta}{2 x} A(x)~.
\label{Bff} 
\eea 
This family of solutions is parameterized by $B$ and the value of the dimensionless quantity 
$\mu R_{\rm NS}$.  
To match continuously the external vacuum dipole solution,  
one must choose $a=\cos(\mu R_{\rm NS})$, $b=\sin(\mu R_{\rm NS})$. 
\begin{figure}[hbt]
   \centering
    \includegraphics[angle=-90,width=0.45\textwidth]{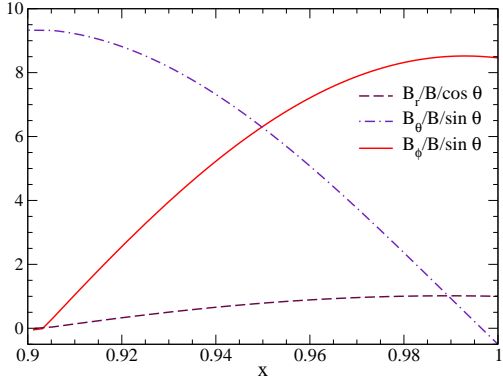}  
\caption{ Normalized magnetic field components for the force free case in the crust:
$B_{r}/B/\cos \theta$ (dashed line), 
$B_{\theta}/B/\sin \theta$ (dashed dotted), and 
$B_{\phi}/B/\sin \theta $ (solid line) vs normalized radial coordinate $x$. 
}
\label{Bfield_FF}
\end{figure}
 
If the magnetic field extends to the center of the NS, only the regular solutions at $x=0$ 
($j_l$) must be considered, i.e., we must set $b=0$, which directly determines $\mu$. 
Due to the superconducting nature of the fluid core, the magnetic field 
may be expelled and confined to the crustal region
\citep{Jones1987,KG2001}. 

This is of course a simplification, since 
in a type II superconductor the magnetic field would be organized in flux tubes with complex 
geometries, but it suffices for our purposes to establish qualitative differences between 
core- and crustal-fields. 
In the case of magnetic fields confined to the crustal region, from the core
radius ($R_{\rm core}$) to $R_{\rm NS}$, 
one must adjust $\mu$ to have a vanishing radial component in the crust-core interface. 
This can be done by solving 
\be 
\tan \left[\mu \left(R_{\rm core} - R_{\rm NS} \right) \right] = \mu R_{\rm core}~. 
\label{root} 
\ee 
The values of the parameter $\mu$ obtained for the NS models used in this paper are listed in 
Table~\ref{NSmasses}. In Fig.~\ref{Bfield_FF} we show the three normalized 
components of the crustal confined force free field for the LM model.

This force-free solution can easily be extended to higher order multipoles, e.g. quadrupole, 
by replacing the angular dependence by the corresponding Legendre polynomial 
and using the corresponding spherical Bessel functions of the same index $l$. From the 
above general expression of force-free fields, some of the cases usually considered in 
the literature can be recovered.
 
\subsubsection{Configurations with other toroidal fields (TC1 and TC2)} 

\begin{figure}[hbt]
   \centering
   \includegraphics[angle=-90,width=0.45\textwidth]{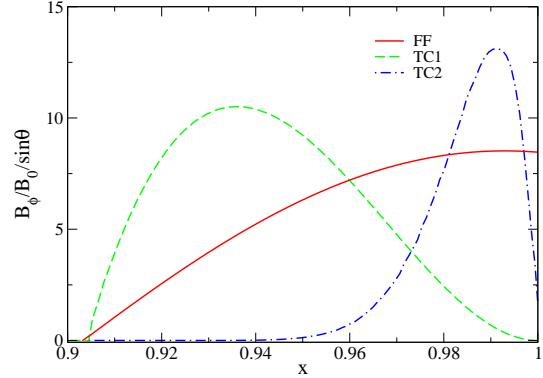}
\caption{ Normalized toroidal field components $B_{\phi}/B/\sin\theta$ vs normalized radial coordinate 
$x$ in the crust. 
Three different models are shown: FF (solid lines), TC1 (dashed lines) and
TC2 (dashed dotted lines). 
}
\label{Bfield_phi}
\end{figure}

We  consider another two 
models with crustal-confined toroidal fields that obey 
\bea
{\cal T}(x,t)&=&{\cal T}_0x(1-x)^2(x-R_{\rm core}/R_{\rm NS}) ~~~~ ({\rm Model~TC1}) \\
{\cal T}(x,t)&=&{\cal T}_0x(1-x)(x-R_{\rm core}/R_{\rm NS})^{10} ~~~ ({\rm Model~TC2})~,
\eea
with the same poloidal component as in the FF case. The constant ${\cal T}_0$ 
is fixed such that $B_{\phi}$ is one order of magnitude larger than $B_r$ at the NS surface. 
In the latter two configurations, the maximum of $B_{\phi}$ is close to the 
crust-core boundary or close to the crust-envelope boundary, respectively. 
The resulting radial profiles of $B_{\phi}$ are shown in
Fig.~\ref{Bfield_phi}. 
We note that the toroidal component of the FF configuration penetrates into
the envelope, while the remaining two (TC1 and TC2) are confined to the crust.

\subsubsection{Crustal poloidal fields (PC model)} 
 
If we assume that the magnetic field is confined to the crust, and maintained by  
purely toroidal currents, we can simply set ${\mathcal T}(x,t)=0$ and, from  
Eq. \ref{bform}, we have 
\bea 
B_r = B \frac{\cos \theta}{x^2} {\mathcal S}(x,t),\quad   
B_{\theta} = - B \frac{\sin \theta}{2 x}\frac{\partial {\mathcal S}(x,t)}{\partial x},\quad  
B_{\phi}= 0, 
\eea 
where again the boundary conditions $\partial {\mathcal S}(x,t)/\partial x =-{\mathcal S}(x,t)$  
at $x=1$, and ${\mathcal S}(x,t)=0$ at $x=R_{\rm core}$ must be fulfilled.
In general, ${\mathcal S}(x,t)$ does not need to coincide with the function $A(x)$ expressed above in terms 
of the spherical Bessel functions. 
However, given the freedom in the choice of ${\mathcal S}(x,t)$, we prefer to use  
the analytical form of $A(x)$ to determine the poloidal field, rather than specifying 
a similar solution that would be equally arbitrary. 
 
\subsubsection{Core dipolar solutions (CD model)} 
The extension of the vacuum solution towards  
the interior can be shown to correspond to the limit $\mu \rightarrow 0$ of the 
non-regular function $n_1$, explicitly, 
\bea 
B_{r} = B\frac{\cos{\theta}}{x^3}, \quad 
B_{\theta} = -B\frac{\sin {\theta}}{2 x^3}, \quad 
B_{\phi}= 0 ~. 
\eea 
Although this solution diverges at $x=0$, it has been used in the literature 
to represent the magnetic field structure in the crust, assuming that the  
field is reorganized in an unknown form in the core. 
Alternatively, one can also take the limit $\mu \rightarrow 0$ of the 
regular spherical Bessel function $j_1$. This leads 
to a homogeneous field aligned with the magnetic axis.

\subsection{Field decay and Joule heating} 
 
 The induction equation that describes the evolution of the magnetic field  
in the crust is 
\bea 
\frac{\partial \vec{B}}{\partial t} &=& -  
\vec{\nabla} \times \left[ \eta \vec{\nabla} \times \vec{B}  
+ \frac{c}{4\pi e n_e} \left( \vec{\nabla} \times \vec{B} \right) \times \vec{B} \right] 
\label{Bdiffusion} 
\eea 
where $\eta=\frac{c^2}{4 \pi \sigma}$ is the electrical resistivity, 
$\sigma$ is the electrical conductivity parallel to the field lines, 
 $n_e$ is the electron density, and $e$ the 
electron charge. 

The first term in the bracket is purely diffusive (Ohmic) and the  
second corresponds to the Hall term.   
Taking the scalar product of $\vec{B}$ by Eq. (\ref{Bdiffusion}), and integrating over 
the volume, it can be seen that the Hall term does not contribute to the dissipation 
of energy, but it redistributes the magnetic energy from one place to another. 
A {\it force-free field} satisfying $\nabla \times \vec{B} = \mu \vec{B}$ is not 
subject to the Hall term and, if $\eta$ is constant throughout the crust volume, 
the induction equation is reduced to 
\bea 
\frac{\partial \vec{B}}{\partial t} = - \eta \mu^2 \vec{B}, \quad 
\eea 
which shows that purely Ohmic dissipation is exponential and proceeds on a typical timescale 
$\tau_{\rm Ohm}=1/\eta \mu^2$.

In a realistic case the situation is more complex, since 
the non-linear evolution of the Hall term must be taken into account 
and the conductivity and electron density profiles are not constant. 
Even if we start from a force-free magnetic field, the effect of a
resistivity gradient leads to a fast modification of its geometry,
and the Hall term becomes immediately important.

For simplicity, and with the purpose of investigating qualitatively the effects  
of magnetic field decay, we include phenomenologically a first stage with rapid 
(non-exponential) decay, and a late stage with purely Ohmic dissipation
(exponential). We assume that the geometry of the field is fixed 
and the temporal dependence is included only in the normalization value 
$B$ according to 
\bea 
B= B_0\frac{\exp{(-t/\tau_{\rm Ohm})}}{1+\frac{\tau_{\rm Ohm}}{\tau_{\rm Hall}}
(1-\exp{(-t/\tau_{\rm Ohm})})} 
\label{Btime} 
\eea 
where $\tau_{\rm Ohm}$ is the Ohmic characteristic time,  
and the typical timescale of the fast, initial stage  
is defined by $\tau_{\rm Hall}$. This is the analytical solution
of the differential equation
\bea
\frac{dB}{dt}= -\frac{B}{\tau_{\rm Ohm}}-\frac{1}{B_0}\frac{B^2}{\tau_{\rm Hall}}
\eea
that takes into account the approximate dependence of the Hall timescale on the magnetic field ($\approx 1/B^2$). We note that $\tau_{\rm Hall}$ should be 
interpreted
as the Hall timescale corresponding to the initial magnetic field strength $B_0$.    
In the early evolution, when $t\ll \tau_{\rm Ohm}$,  
\bea 
B \simeq  B_0 (1+t/\tau_{{\rm Hall}})^{-1} 
\label{BHall}
\eea 
while for late stages, when $t \geq \tau_{\rm Ohm}$ 
\bea 
B \simeq  B_0 \exp(-t/\tau_{{\rm Ohm}}) 
\label{BOhm}
\eea 
 
This simple law reproduces qualitatively the results from more complex simulations 
\citep{PonsGeppert2007} and facilitates the implementation of field decay 
in the cooling process of NSs for different 
Ohmic and Hall timescales, treated as simple constant parameters. The initial 
{\it Hall} stage, in which the Hall drift qualitatively affects the thermal evolution, 
is of particular importance for models of highly magnetized NS, e.g, for magnetars the field can
dissipate 75\% of the energy in $ \approx\tau_{\rm Hall}$. In contrast, the late {\it Ohmic} 
stage lasts for about $\tau_{{\rm Ohm}}\simeq 10^6$ yr.

If the field is anchored into the superconducting core, 
the results will be different. It is not the
purpose of this paper to discuss such a possibility, which deserves
a separate analysis, but to investigate the
possible effects of crustal fields that enhance the surface temperature anisotropy
and are subject to Ohmic dissipation and, consequently, Joule heating.

\section{Thermal evolution} 
 \label{SecThermal}
\subsection{The diffusion equation in axial symmetry} 

Assuming that deformations with respect to the spherically-symmetric case due 
to rotation, magnetic field, and temperature distribution do not affect 
the metric in the interior of a NS, we use the standard form \citep{Misner1973}
\bea
ds^2= -e^{2\Phi}dt^2+ e^{2\Lambda}dr^2 + r^2 d\Omega^2~.
\label{metric}
\eea
Using this background metric but considering an axially-symmetric
temperature distribution,   
the thermal evolution of a NS can be described by the energy balance 
equation 
\bea 
c_{v} e^{\Phi} \frac{\partial T}{\partial t} + \vec{\nabla} \cdot  
(e^{2\Phi } \vec{F})= 
e^{2 \Phi} Q  
\label{eneq} 
\eea 
where $c_v$ is the specific heat per unit volume and $Q$ is the  energy loss/gain  
 by $\nu$-emission, Joule heating, accretion heating, etc.. 
In the diffusion limit, the heat flux is simply 
\bea 
\vec{F} = -e^{-\Phi } \hat{\kappa} \cdot \vec{\nabla} (e^{\Phi } T) 
\label{fluxeq} 
\eea 
where $\hat \kappa$ is the thermal conductivity tensor.  
Defining the redshifted temperature to be $\tilde{T} \equiv e^{\Phi } T$,  
the  components of the redshifted flux $\tilde{F} \equiv e^{2\Phi } F$ can be written explicitly as follows 
\bea 
\tilde{F}_{r} = - e^{\Phi }\left(\kappa_{rr} e^{-\Lambda} \partial_{r} \tilde{T} +  
               \frac{\kappa_{r \theta}}{r} \partial_{\theta} \tilde{T}\right) \nonumber \\ 
\tilde{F}_{\theta} = - e^{\Phi} \left(\kappa_{\theta r} e^{-\Lambda} \partial_{r} \tilde{T} +  
              \frac{\kappa_{\theta \theta}}{r} \partial_{\theta} \tilde{T} \right) 
\eea 
where the $\phi$-component is not relevant because of the axial symmetry. 
 
The total conductivity tensor, $\hat{\kappa}$,  must include the contributions 
of all relevant carriers, which are of interest in the solid crust: electrons, neutrons, protons and 
phonons 
\bea
\hat{\kappa} = \hat{\kappa}_e + \hat{\kappa}_n + \hat{\kappa}_p + \hat{\kappa}_{ph}~.
\eea

The heat is transported primarily by electrons, which provide the dominant
contribution. Radiative transport is important close to the surface, but the outer
region is considered by means of boundary conditions (Sect.~\ref{BConditions}) 
in place of direct calculation.

For magnetized NS, the electron thermal conductivity tensor becomes anisotropic: 
in the direction perpendicular to the 
magnetic field,  its strength is strongly diminished, which causes a suppression of the heat flow  
orthogonal to the magnetic field lines.  
The ratio of conductivities along and orthogonal to the magnetic field can be defined in terms of 
the magnetization parameter, $\omega_B \tau$, as
\bea 
\frac{\kappa^{\parallel}_e}{\kappa^{\perp}_e} = 1 + (\omega_{B} \tau)^{2}~,
\label{Magparameter}
\eea 
where $\tau$ is the electron relaxation time \citep{Urpin1980}, and
$\omega_B $ is  the classical electron gyrofrequency 
corresponding to a magnetic field strength $B$ 
\bea
\omega_B = \frac{eB}{m_e^* c}~,
\label{efreq}
\eea
where $m_e^*$ is the electron effective mass.
The dimensionless quantity $\omega_B \tau$ is an indicator of the 
suppression of the  thermal conductivity
in the transverse direction. When $\omega_B\tau \gg 1$, the effects of the magnetic field 
on the transport properties are crucial.

In spherical coordinates, and choosing the polar axis 
to coincide with the axis of symmetry of the magnetic field,  
the electron contribution can be written as follows 
\bea 
\hat{\kappa}_e& =& \kappa^{\perp}_e
\left( \hat{I}+(\omega_{B} \tau)^2 \left( \begin{array}{ccc}  
 b_{rr}  & b_{r\theta} & b_{r\phi} \\ 
 b_{r\theta}  & b_{\theta\theta} & b_{\theta\phi} \\ 
 b_{r\phi}  & b_{\theta\phi} & b_{\phi\phi} \\ 
               \end{array} \right) +  
  \omega_{B}\tau \left(  \begin{array}{ccc} 
      0      & b_{\phi} & -b_{\theta} \\ 
  -b_{\phi}  &    0     &  b_{r}      \\ 
   b_{\theta}& -b_{r}   &   0    \end{array}  
\right) \right) ~, 
\nonumber \\  
\eea 
where $\hat{I}$ is the identity matrix, and $b_{r}, b_{\theta}, b_{\phi}$ are  
the components of the unit vector in the direction of the magnetic field, and 
$b_{ij}=b_ib_j$ for $i,j=r,\theta, \phi$. 
Using the above expression for $\hat{\kappa}_e$, the electron part of the flux reads, in closed form: 
\bea 
\vec{F}_e&\!=\!& 
-e^{\Phi}\kappa^{\perp}_e  
\left[ \vec{\nabla} \tilde{T} + \left(\omega_{B} \tau \right)^{2} \left( \vec{b}  
\cdot \vec{\nabla} \tilde{T} \right)  
\cdot \vec{b} + \omega_{B} \tau \left( \vec{b} \times \vec{\nabla} \tilde{T} \right) \right]~.
\label{flux} 
\eea 
 
The thermal evolution Eq.~(\ref{eneq}), with the above expression for 
the fluxes, is solved numerically for a given background magnetic field with fixed 
geometry and strength that varies with time according to Eq.~(\ref{Btime}). 
The emissivity terms on the right-hand side of Eq.~(\ref{eneq}) and the specific heat of the first term of the same equation 
are considered in the next section.
 
\subsection{Boundary conditions} 
\label{BConditions}

For numerical reasons, the thermal evolution equation is difficult to solve in 
the thin layer which consist of the envelope, of a few meters depth,  and the atmosphere, of a few centimeters,  in which radiative equilibrium is established and the observed spectrum 
is generated. Since this outer layer has a small scale and its thermal relaxation time 
is much shorter than the overall evolutionary time, the usual approach is to use 
results of stationary, plane-parallel, envelope models to obtain a phenomenological fit 
that relates the temperature at the bottom of the envelope $T_b$, with the surface 
temperature $T_s$. This ``$T_b$--$T_s$ relationship'' can be used to implement boundary 
conditions, because the surface flux can then be calculated for a given temperature at the base of the 
envelope, which corresponds to the outer point of the numerical grid in our cooling simulations. 
 
Models assuming a non-magnetized envelope made of iron and iron-like nuclei  
show that the surface temperature is related to $T_b$ as follows \citep{Gudmundsson1983}  
\be 
T_{b,8}
=1.288 
\left[ \frac{T_{s,6}^4}{g_{14}}
\right]^{\,0.455} 
\ee 
where $g_{14}$ is the surface gravity in units of $10^{14}$ cm~s$^{-2}$, $T_{b,8}$ is $T_b$ in $10^8$~K, and  
 $T_{s,6}$ is $T_s$ in $10^6$~K.
 
Since the magnetic field increases the heat permeability of the envelope in  
regions where the magnetic field lines are radial but strongly suppresses it where 
the magnetic field lines are almost 
tangential \citep{Potekhin2001}, this implies a large anisotropic distribution of $T_s$, 
which depends on the magnetic field geometry. 
In iron magnetized envelopes, 
the surface temperature depends on the angle $\varphi$ that the magnetic field forms  
with respect to the normal to the NS surface by means of a function $\mathcal{X}$:   
\be 
   T_\mathrm{s}(B,\varphi,g,T_b)\approx 
   T_\mathrm{s}^{(0)}(g,T_b) \,\mathcal{X}(B,\varphi,T_b) ,  
\label{PCY-iron} 
\ee 
where  
\begin{equation} 
   T_\mathrm{s}^{(0)} \approx  10^6 \, 
   g_{14}^{1/4}\left[(7\zeta)^{2.25}+(\zeta/3)^{1.25}\right]^{1/4}~~{\rm K}, 
\end{equation} 
and 
$\zeta\equiv 0.1 T_{b,8} -0.001\,g_{14}^{1/4}\,\sqrt{0.7 \, T_{b,8}}$.
The function $\mathcal{X}$ was fitted by  
decomposing into transversal and longitudinal parts as  
\bea 
 \mathcal{X}(B,\varphi,T_b) &=&  
      \big[ \,\mathcal{X}_\|^{9/2}(B,T_b)  \cos^2\varphi
\nonumber 
   + \,\mathcal{X}_\perp^{9/2}(B,T_b) \sin^2\varphi \big]^{2/9}~, 
 \label{fit3} 
\eea 
which is valid for $B < 10^{16}$~G and  
$10^7 \mbox{ K} \leq T_b \leq 10^{9.5}$~K, with the additional constraint that
$T_\mathrm{s}>2\times10^5$~K. 
 
Strongly magnetized envelopes were revisited by \cite{Potekhin2007}, who reconsidered
neutrino emission processes that are activated by strong fields, that is neutrino synchrotron. 
These processes
were found to lower the surface temperature at a fixed $T_b$. To take this effect into account, 
we introduce a maximum surface temperature $T^{\rm max}_s(\varphi)$ that can be reached for a given 
$T_b$, which  we parameterize as a function of $B$ to reproduce their results.
 
In this work, we assume an iron composition for the envelope and focus on the 
magnetic corrections to the transport due to the presence  of a large field. 
Nevertheless, if light elements were present, which is 
very unlikely because the large magnetic field suppresses accretion, 
they strongly reduce the blanketing effect and the relations used here should be revised.  
Another possibility is that the gaseous atmosphere and the outer envelope condensates
to a solid state due to the cohesive interaction between ions caused by the magnetic
field. This condensed surface has different emission properties and, consequently, 
the boundary condition must be recalculated, as for example in \cite{Azorin2006}.
This scenario will be studied in future work.

\section{Microphysics inputs}
\label{SecMicrophysics}
\subsection{EoS and superfluidity}
\label{SFsection}

To build the background NS model, we used a  
Skyrme-type equation of state (EoS), at zero temperature to
describe both the NS crust and the liquid core, based on  
the effective nuclear interaction SLy \citep{Douchin2001}. 
The low density EoS, below the neutron drip point, employed is 
that of Baym et al. (1971). 
Throughout this work we use two models:
a low mass neutron star (LM) and a high mass (HM) NS,
which have the properties listed in Table~\ref{NSmasses}. 
\begin{table}
\begin{minipage}[t]{\columnwidth}
\caption{ Central density $\rho_c$, mass $M$, radius $R_{\rm NS}$,
crust thickness $\Delta R_{\rm crust}$, and $\mu$ parameter for the two cases used 
in this work: low mass (LM) and 
high mass (HM). }
\begin{tabular}{c|c|c|c|c|c}
 Model &$\rho_c$ &  $M$  & $R_{\rm NS}$ & $\Delta R_{\rm crust}$ & $\mu$ \\
       &(g~cm$^{-3}$) &  ($M_{\odot}$) & (km) & (km) & (km$^{-1}$) \\
\noalign{\smallskip}\hline
LM &$8.1\,10^{14} $&  $1.35$& $12.83$& $1.24$& $1.32$\\
HM &$1.1\,10^{15} $&  $1.63$& $12.36$& $0.86$& $1.87$\\
\end{tabular}
\end{minipage}
\label{NSmasses}
\end{table}
For the chosen EoS, the crust-core interface is at 
$0.46\, \rho_0$, where $\rho_0=2.8 \times 10^{14}$ g~cm$^{-3}$ is the nuclear saturation density, 
and for both models the crust thickness is approximately $1$ km, 
which defines a characteristic length scale for the confinement of the crustal magnetic field. 

\begin{figure}[bth]
   \centering
   \includegraphics[angle=-90,width=0.5\textwidth]{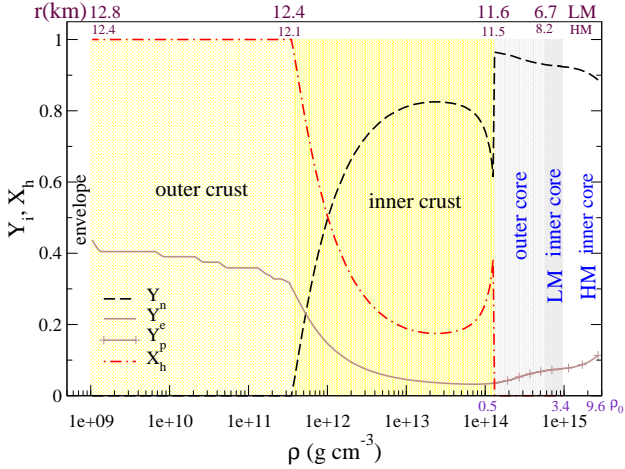}
\caption{NS composition for the EoS employed. Number of particles per baryon as a function of the density: $Y_{e}$ (solid line), $Y_p$ (solid line with plus symbols), and $Y_n$ (dashed line). $X_h$ is indicated by 
dashed-dotted lines.  
The scale in the upper horizontal axis indicates the corresponding 
radial coordinate at each density for both LM and HM models. }
\label{NScomposition}
\end{figure}

In Fig.~\ref{NScomposition}, the number of particles per baryon ($Y_{(n,p,e)}$)  
and the fraction of nucleons
inside heavy nuclei ($X_h$) are shown as a function of the density.  
In the upper horizontal axis, the 
scale shows the value of the radial coordinate that limits each region: 
the outer and inner crust, and the outer and inner core, 
for the LM and HM NS models. 
 We do not include muons in our equation of state.  

\begin{figure}[bth]
   \centering
   \includegraphics[angle=-90,width=0.45\textwidth]{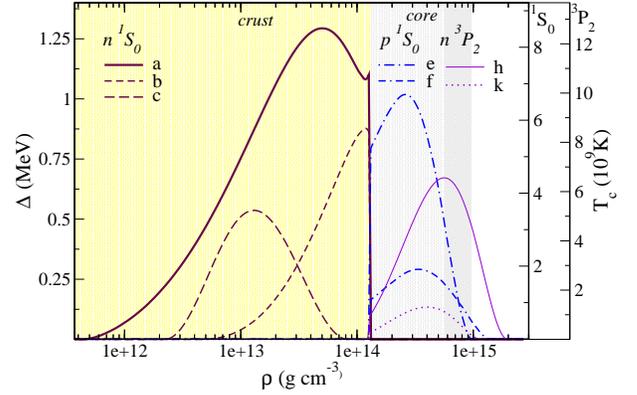}
\caption{Energy gaps for superfluidity as a function of the density. 
Lines denote: case $a$ (thick solid), case $b$ (short dashed), 
and case $c$ (long dashed) for  $n$$^1S_0$; case $e$ (dashed dotted) and
case $f$ (double dashed dotted) for  $p$$^1S_0$; 
case $h$ (thin solid) and case $k$ (dotted) for $n$$^3P_2$. 
The right axes show $T_c$ for $^1S_0$ and $^3P_2$ pairing states.}
\label{FigGaps}
\end{figure}

Pairing in nuclear matter can play an important role 
in NS cooling, without affecting significantly the EoS, but strongly modifying 
neutrino emissivities and specific heat. 
In fact, for the paired component, these are suppressed by exponential Boltzmann 
factors.  
If superfluidity (SF) occurs inside NSs, i.e. when $T$ is below a critical temperature 
($T_c$),
the inclusion of these suppression factors will have important 
consequences on the thermal evolution as we see in the following subsections. 
We consider the pairing of  neutrons in the crust in the $n$~$^1S_0$ state, 
protons  in the core in the $p$~$^1S_0$ state,  and the $n$~$^3P_2$ state, for neutrons in
the core. 
Following \cite{Kaminker2001} and \cite{Andersson2005}, 
we use a phenomenological formula for the momentum dependence of the energy gap at zero temperature
\be
\Delta(k_{F,N})=\Delta_0 \frac{(k_{F,N}-k_0)^2}{(k_{F,N}-k_0)^2+k_1^2}\frac{(k_{F,N}-k_2)^2}
{(k_{F,N}-k_2)^2+k_3^2}
\label{Gapsparam}
\ee
where $k_{F,N}=(3\pi^2n_N)^{1/3}$ is the Fermi momentum and $n_N$ is the particle
density of each type of nucleons ($N=n,p$) involved.
The parameters $\Delta_0$ and $k_i$, $i=1..4$ are values fitted 
to recent model calculations listed 
in Table~\ref{Tablegaps}. This expression 
is valid for $k_0<k_{F,N}<k_2$, with vanishing $\Delta$ outside this range. The density
dependence of the gaps is plotted in Fig.~\ref{FigGaps}.
\begin{table}
\begin{minipage}[t]{\columnwidth}
\caption{Parameterization and references of the energy gaps for superfluid states}
\begin{tabular}{c|rrrrr|c}
Label&$\Delta_0$ ~~ &$k_0$~~~ &$k_1$~~~&$k_2$~~~&$k_3$~~~&Ref. \\
     & (MeV) &(fm$^{-1}$)&(fm$^{-1}$)&(fm$^{-1}$)&(fm$^{-1}$)&\\
\hline\noalign{\smallskip}
\multicolumn{7}{l}{$n$~$^1S_0$}\\
\hline
$a$ &68& 0.1&   4& 1.7&   4& 1\\
$b$ & 4& 0.4& 1.5&1.65&0.05& 2\\
$c$ &22& 0.3&0.09&1.05&   4&  3\\
\hline\noalign{\smallskip}
\multicolumn{7}{l}{$p$~$^1S_0$}\\
\hline
$e$ & 61&   0&  6&  1.1& 0.6& 4  \\
$f$ & 55&0.15&  4& 1.27&   4& 5 \\
\hline\noalign{\smallskip}
\multicolumn{7}{l}{$n$~$^3P_2$}\\
\hline
$h$ &  4.8&  1.07& 1.8& 3.2&   2& 6 \\
$k$ & 0.42&  1.1 & 0.5& 2.7& 0.5&  7\\
$m$ & 2.9 &  1.21& 0.5& 1.62& 0.5 &4\\
\noalign{\smallskip}
\end{tabular}

References. (1)~\cite{Wambach1993}; (2)~\cite{Schulze1998}; (3)~\cite{Chen1986}; 
(4)~\cite{Elgaroy1996}; (5)~\cite{Amundsen1985}; (6)~\cite{Baldo1998}; (7)~\cite{Elgaroy1996a}
\label{Tablegaps}
\end{minipage}
\end{table}

For the $n$ $^1S_0$ pairing, the bare interaction predicts a maximum gap 
$\Delta^{\rm max} \simeq 3$ MeV \citep{Schulze1998}, 
but the polarization effects reduce it by a factor $2$-$3$, giving 
$\Delta^{\rm max} \simeq 1$ MeV at  $k_{F,n}\simeq 0.7-0.8$ fm$^{-1}$. 
This is the case for 
the approximation $a$, which in our NS models peaks at $\rho =4 \times 10^{13}$ g~cm$^{-3}$ 
in the inner crust (Fig.~\ref{FigGaps}). 
Although model calculations have found some agreement
about the value of the maximum energy gaps, 
its precise location is uncertain and may vary in 
the different approaches, as shown in cases $b$ and $c$. 
The corresponding critical temperatures for the $s$-wave 
can be calculated approximately to be $T_c=0.56\,\Delta(T=0)$, 
which implies a  maximum for neutrons of $T^{\rm max}_{c,n}=9 \times 10^9$ K, 
for the case $a$. As shown later, this high temperature implies
that neutrons become superfluid in the crust during the 
early cooling of a NS and the most important consequence is that the crustal 
specific heat is suppressed. 

The calculations for the $p$~$^1S_0$ pairing take into account the presence 
of the neutron gas and depend also on the proton fraction through 
the symmetry energy of the EoS. For different approaches, such as case $e$ and $f$, 
$\Delta^{\rm max}$ is located at about $k_{F,p}\simeq 0.4-0.5$ fm$^{-1}$, 
which is much smaller than for neutrons, due to the smaller proton effective mass. 
Nevertheless, due to the proton number density, the peak is shifted to 
$\rho \simeq 2 \times10^{14}$ g~cm$^{-3}$  
in the outer core of our NSs, as shown in Fig.~\ref{FigGaps}. Most models agree that the proton 
energy gap should vanish at $k_{F,p}> 1.5$ fm$^{-1}$, i.e. at high densities inside the star
$\rho \gtrsim 10^{15}$ g~cm$^{-3}$.
For the cases considered here, $T^{\rm max}_c \simeq 2$-$6 \times 10^9$ K, indicating 
that also proton superfluidity
will be present from the very beginning of the NS thermal evolution. 
Due to the charge of the protons, 
the superfluid is also in a superconducting (SC) state.

The situation for the pairing of neutrons in the core is more complicated, 
because the $^3P_2$ state has coupled anisotropic gap equations. 
Some calculations show that the energy gap should be reduced by a factor
of $2$-$3$, as in the proton case, due to the lower neutron effective mass in very dense matter. 
But relativistic effects become important and there is no conclusive approach to the problem. 
Thus, in our calculations we considered three different cases that reflect this uncertainty: 
$h$, $k$ and $m$ with 
$\Delta^{\rm max} \simeq 0.6, 0.1, 0.02$ MeV, respectively (Table~\ref{Tablegaps}). 
The location of the maximum varies as well, at $k_{F,n} \simeq 1.4$-$2$ fm$^{-1}$, i.e. 
at $\rho \simeq 2$-$6\times 10^{15}$ g~cm$^{-3}$,
as plotted in Fig.~\ref{FigGaps}, in which case $m$ is omitted because it is not visible in this scale. 
We note that for the $p$-wave we take that $T_c=0.82\,\Delta(T=0)$ 
\citep{BL84},
which corresponds to a a wide range of 
$T^{\rm max}_{c,n} \simeq 2 \times 10^8$-$6 \times 10^9$ K for the chosen models. 

The temperature dependence of the energy gap that we use is the approximate 
functional form given by \cite{Levenfish1994}:
\be
\frac{\Delta(T)}{kT} \approx \sqrt{1- \frac{T}{T_c}}
~\left(\alpha - \frac{\beta}{\sqrt{T/T_c}} 
+\frac{\gamma}{T/T_c}\right)
\label{Temperaturegaps}
\ee
where $\alpha=1.456$, $\beta=0.157$, and $\gamma=1.764$  for $^1S_0$, 
and $\alpha=0.789$, $\beta=0$, and $\gamma=1.188$  for $^3P_2$ states. 

From these considerations, it is clear that a NS at the  
beginning of its cooling history should contain superfluid neutrons in 
the crust and superconducting protons in the core, while  
the occurrence of neutron pairing in the core is rather model dependent. 

\subsection{Thermal conductivity}

In NS cooling simulations, the thermal conductivity should be calculated over a
region covering a large range of densities,
from the core ($\approx 10^{15}$ g~cm$^{-3}$) to the outer crust
($\approx 10^{9}$ g~cm$^{-3}$).

Schematically, the thermal conductivity tensor can be written for each carrier 
in terms of the effective relaxation time tensor, ${\hat \tau}_{\rm eff}$
\citep{Flowers1976,Urpin1980,Itoh84},
as follows,
\bea
\hat{\kappa} &=&  \frac{\pi^{2} k_{B}^{2} n c^{2} T}{3 m^*} {\hat \tau}_{\rm eff},
\eea
where $n$ is the carrier number density, $m^*$ is its effective mass, and 
$\hat{\tau}_{\rm eff}$ is a tensor whose components are interpreted as effective 
relaxation times. In the non-quantizing case, these relaxation times can be written 
in terms of the non-magnetic relaxation time, which is calculated to be the inverse
of the sum of all collision frequencies of the processes involved. 

\begin{figure}[hbt]
   \centering
   \includegraphics[angle=-90,width=0.5\textwidth]{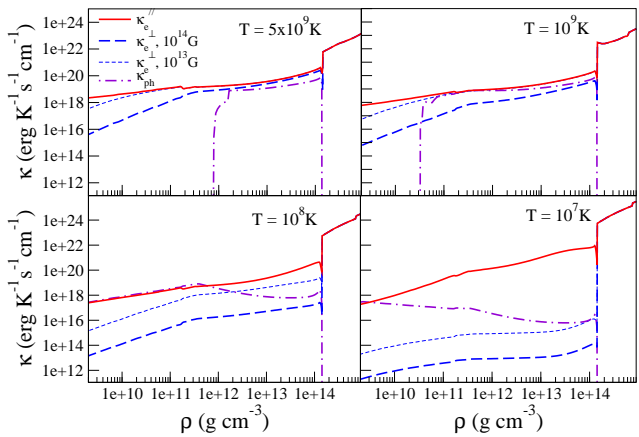}
\caption{Thermal conductivity contributions as a function of the density for different fixed temperatures.   
$\kappa_e^{\parallel}$ is shown with solid lines, $\kappa_e^{\perp}$ 
 with dashed lines (thick for $B=10^{13}$ G and thin for  $B=10^{14}$ G),
 and $\kappa_{ph}$ with dotted dashed lines. }
\label{Kthermal}
\end{figure}

In the inner liquid core, we include contributions from electrons, 
neutrons and protons \citep{Gnedin1995,Baiko2001}, 
without taking account of the effects of the magnetic field because
of proton superconductivity: the field is either expelled from
the core or confined into flux tubes that occupy a small fraction
of its volume.
We note that, if the magnetic field does not affect transport properties,
a large thermal conductivity of matter is produced soon 
after birth in an isothermal core (Fig.~\ref{Kthermal}), 
which implies that the precise value of the thermal conductivity is not important.

In the solid crust, only electron and phonon transport are considered.
While phonon conductivity is negligible in non-magnetic neutron stars,
this situation changes when the magnetization parameter becomes large. 
Since electron transport is drastically suppressed in the direction transverse 
to the magnetic field, 
the phonon contribution may become dominant at low density as shown in Fig.~\ref{Kthermal}. 

In our calculations, we use the non-quantizing electron conductivities from the public code 
of A. Potekhin (1999)\footnote{{\tt www.ioffe.rssi.ru/astro/conduct/condmag.html}}.
The three electron scattering processes that play a role in our scenario are scattering
off ions, electron-phonon scattering, and scattering  off impurities.
Semi-analytic expressions and fitting formulae for the relaxation time and thermal
conductivity along the magnetic field for all three processes, were derived
by Potekhin \& Yakovlev (1996).

At high temperatures, the phonon conductivity of the lattice is determined mainly by
Umklapp processes, and can be approximated by the expression
\bea
\kappa_{ph} = \frac{1}{3}{c_v c_s \lambda_{ph}} 
\eea
where $c_s$ is the sound speed, $c_v$ the specific heat, and
$\lambda_{ph}$ the phonon mean free path in the lattice. In Fig.~\ref{Kthermal}, it can be seen 
that the phonon 
contribution becomes more important at lower densities as the temperature 
decreases and the liquid solidifies 
into a lattice. 

Chugunov \& Haensel \cite{CH2007} revised the ion
thermal conductivity in neutron star envelopes. They included the contribution of
electron-phonon scattering and improved the calculations of phonon-phonon scattering.
Our estimates for $\lambda_{ph}$ are larger than their results by a factor of a few,
depending on the density, which results in a smaller temperature anisotropy.
However their results are more applicable to the neutron star envelope, 
than for the crust. The main reason is that, 
at temperatures smaller than the Debye temperature, 
the inclusion of the effect of impurities and defects in the crystal becomes necessary. 
Given our limited knowledge of the impurity content of the inner crust, which
may affect the results, we do not include phonon-impurity interactions in our
simulations. In principle, its effect would be to reduce the phonon mean free path, 
but it is unclear how to calculate accurately this contribution at low temperature.

\subsection{Specific heat}
\label{SHeatsection}
\begin{figure}[hbt]
   \centering
   \includegraphics[angle=-90,width=0.5\textwidth]{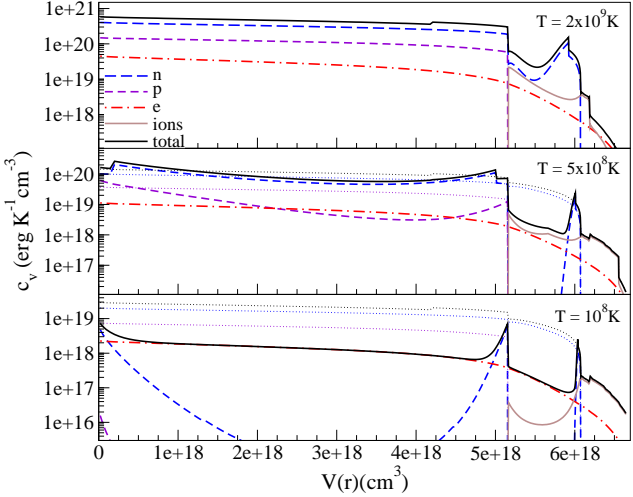}
\caption{Specific heat 
as a function of the enclosed volume $V(r)$ within a radius $r$, 
and for different fixed $T$. We show the contribution of 
neutrons (long-dashed line), 
protons (short-dashed line), electrons (dotted-dashed line), and ions 
(brown solid line). Total $c_{v}$ is plotted using a black solid line. 
The dotted lines show the non-superfluid case. 
}
\label{FigCv}
\end{figure}

In normal non-superfluid neutron star matter,
most of the total heat capacity of a NS star originates in the nucleons in the core. 
For degenerate fermions of type $i$, the specific heat per unit volume in terms of the dimensionless  
Fermi momentum $x_{F,i}=\hbar k_{F,i}/m_i c$ is
\bea
c_{v,i}=\pi^2 \frac{n_i k^2 T}{m_i c^2}\frac{(x_{F,i}^2+1)^{1/2}}{x_{F,i}^2}.
\eea
Then, the contribution of relativistic electrons is
\bea
c_{v,e}\simeq 5.4\,10^{19}\, \left(\frac {n_e}{n_0}\right)^{2/3} T_9 
\quad{\rm erg~cm}^{-3}{\rm K}^{-1}
\eea
while for non-relativistic nucleons $N=n,p$  is
\bea
c_{v,N}\simeq 1.6\,10^{20}\,\frac{m^{\ast}_N}{m_N} 
\left(\frac {n_N}{n_0}\right)^{1/3} T_9 \,\mathcal R^{cv}
\quad{\rm erg~cm}^{-3}{\rm K}^{-1}~,
\eea
where $n_0=0.16$ fm$^{-3}$.
We include the effect of superfluidity through the factor $\mathcal R^{cv}$ 
\citep{Levenfish1994}, which depends on the pairing state of the 
nucleons involved ($^1S_0$ or $^3P_2$).
The electron contribution, or that of muons, if present,  inside the core is, in principle, much smaller,
but it dominantes when all nucleon species undergo a phase transition to a superfluid 
state  (see Fig.~\ref{FigCv}).

In our model we include the crustal specific heat, which has contributions from the neutron gas, 
the degenerate electron gas and the nuclear lattice \citep{VanRiper1991}; it is however negligible
in comparison to the core contribution, due to the small volume of the crust.  
\subsection{Neutrino emissivities}
\label{Neutrinosection}

During the first  $\approx 10^5$ yr,  the so-called neutrino cooling era,
the evolution of a NS is governed by the 
emission of neutrinos. Thereafter, photons 
radiated from the surface control the evolution  in the photon cooling era.
The path of a NS in a temperature-age diagram and the duration of the neutrino
cooling era is determined by the efficiency of the neutrino processes in their interior. 
Typically, neutrino emissivities, at high densities, depend on 
temperature to the power of a large number, 
and weakly on density;  a review is provided by \cite{YakovReport2001}. 
In the so-called {\it standard cooling} scenario, the total emissivity is dominated by 
slow processes in the core, such as modified Urca (MUrca) and nucleon--nucleon ($N$-$N$) Bremsstrahlung.
The {\it minimal cooling} model, in which pairing between nucleons 
and the effects of superfluidity are both included \citep{Page2004}, is more realistic. 
If fast neutrino processes, i.e. direct Urca (DUrca), occur, the evolution of a NS alters significantly, 
leading to the {\it enhanced cooling} scenario.
Nevertheless, DUrca only operates above a critical proton fraction $Y_p^c \approx 0.11$,
that is only reached at high density ($4$--$6$) $\rho_0$ in the inner core of high mass NSs.  
Since we assume a superfluid core with $^1S_0$ paring for protons and $^3P_2$ pairing for neutrons, 
we account for the exponential suppression of these processes through reduction functions
$\mathcal R$ \citep{YakovReport2001}. 
 We  include the Cooper Pair Breaking and Formation emissivity (CPBF), 
although the effectivity of this process was questioned both, 
by observational  arguments \citep{Cumming2006} and 
by theoretical calculations \citep{Leinson2006}. In the latter work, the authors showed that
the neutron  $^1S_0$ CPBF emissivity is suppressed, which is relevant in the crust, 
but the proton  $^1S_0$ and neutron  $^3P_2$ channels, which are relevant in the core, 
are not seriously altered.
Including or not this suppression has an effect on the
early relaxation of the crust, but has little imprint on the long term cooling evolution.

In our calculations, we consider all relevant neutrino emission processes listed in Table 
\ref{Neutrinocore}, which indicates the density and temperature dependence of the emissivity
for the different processes.
The factors that account for further corrections, due to for example effective masses and correlation effects, 
can be found in the references listed in Table~\ref{Neutrinocore}. 
The table also includes the critical proton fraction $Y_p^c$ 
that is required before some processes can operate.
In the {\it enhanced cooling} scenario, 
we include the fast DUrca process. The efficiency of this fast reaction 
is exponentially reduced when superfluidity is taken into account. 
\begin{table}[htb] 
\begin{minipage}[t]{\columnwidth}
\caption{Neutrino processes and their emissivities $Q$ 
in the core and in the crust. 
Third column shows the onset for some processes to operate 
(critical proton fraction $Y_p^c$). 
Detailed functions and precise factors can be found in the references 
(last column). 
}

\begin{tabular}{l|l|l|c}
Process &$Q\,[{\rm erg\, cm^{-3}s^{-1}}]$ 
                    &Onset&Ref\\ 
\hline\noalign{\smallskip}
\multicolumn{4}{l}{Processes in the core}\\ 
\noalign{\smallskip}
\hline
MUrca ($n$-branch)&&&\\
 $nn\rightarrow pne\bar\nu_e$ &&&\\
 $pne\rightarrow nn\nu_e$ 
 & 
$8\times 10^{21}\,\mathcal R^{MU}_n \,n_p^{1/3} 
\,T^8_9$ &&1  \\ 
MUrca ($p$-branch) &  & &\\ 
 $np\rightarrow ppe\bar\nu_e$ &&&\\
 $ppe\rightarrow np\nu_e$  
 & 
$8\times 10^{21} \,\mathcal R^{MU}_p \,n_p^{1/3} 
 \,T^8_9$ &$Y_p^c=0.01$  & 1 \\ 
\hline
NN-Bremsstrahlung  & &&\\ 
$nn\rightarrow nn \nu \bar\nu$ & 
$7\times 10^{19}\, \mathcal R^{nn} \,n_n^{1/3}
\,T^8_9$ && 1 \\ 
$np\rightarrow np \nu \bar\nu$ &
$ 1\times10^{20} \,\mathcal R^{np} \,n_p^{1/3}
\,T^8_9$  & & 1 \\ 
$pp\rightarrow pp \nu \bar\nu $ &
$7 \times 10^{19}\,\mathcal R^{pp} \,n_p^{1/3}
\,T^8_9$  & & 1 \\ 
\hline
$e$-$p$ Bremsstrahlung   & &&\\ 
$ep\rightarrow ep \nu \bar\nu $ &
$2\times 10^{17}\,n_B^{-2/3}
\,T^8_9$  && 2 \\
\hline
DUrca &&&\\
$n\rightarrow pe\bar\nu_e, pe \rightarrow n\nu_e$& 
$4\times 10^{27}\,\mathcal R^{DU} \,n_e^{1/3}
\,T_9^6$
&$Y_p^c=0.11$& 3\\ 
$n\rightarrow p\mu\bar\nu_{\mu},p\mu \rightarrow n\nu_{\mu}$ 
& $4 \times 10^{27}\,\mathcal R^{DU} \,n_e^{1/3}
\,T_9^6$ 
&$Y_p^c=0.14$& 3\\
\hline\noalign{\smallskip}
\multicolumn{4}{l}{Processes in the crust}\\
\noalign{\smallskip}
\hline
Pair annihilation  & & &\\ 
$ee^+\rightarrow \nu \bar\nu$ &
$9 \times 10^{20}\, F_{\rm pair}(n_e,n_{e^+})$ &  & 4\\
&&\\
\hline
Plasmon decay &&&\\
$\tilde e\rightarrow \tilde e\nu\bar\nu$ &
$1\times 10^{20} \, I_{\rm pl} (T,y_e)$
& & 5\\
\hline
$e$-$A$ Bremsstrahlung &&&\\
$e (A,Z)  \rightarrow e (A,Z) \nu \bar\nu$&
$3 \times 10^{12}\, L_{eA}\,Z\,\rho_o\,n_e\,T^6_9$
& &6\\
\hline 
$N$-$N$-Bremsstrahlung   &&&\\ 
$nn\rightarrow nn \nu \bar\nu$ & 
$7 \times 10^{19}\, \mathcal R^{nn}f_{\nu} \,n_n^{1/3}
\,T^8_9$ & & 1  \\ 
\hline\noalign{\smallskip}
\multicolumn{4}{l}{Processes in the core and in the crust}\\ 
\noalign{\smallskip}
\hline
CPBF && &\\
$\tilde B + \tilde B \rightarrow \nu \bar \nu$
&$1\times 10^{21}\,n_N^{1/3} 
\,F_{A,B}\, T^7_9$&
& 7\\
\hline 
Neutrino synchrotron&&&\\  
 $e \rightarrow (B) \rightarrow e \nu \bar\nu$ 
&$9\times10^{14}\,S_{AB,BC}
\,B_{13}^2 \,T^5_9$ &  
& 8\\
  \end{tabular} 
  
\vspace{0.3cm}
Ref. (1)~\cite{YakovlevLevenfish1995}; (2)~\cite{Maxwell1979}; (3)~\cite{Lattimer1991}; 
(4)~\cite{Kaminker1994}; (5)~\cite{YakovReport2001}; (6)~\cite{Haensel1996,Kaminker1999}; 
(7)~\cite{Yakovlev1999A}; (8)~\cite{Bez1997}
 
 \label{Neutrinocore} 
\end{minipage}
\end{table} 

The neutrino energy losses from processes that occur inside the crust are 
very important at the beginning 
of thermal evolution, during the relaxation stage prior to the core-crust thermal coupling. 
This stage lasts about $10-10^2$ yr and was studied in detail by \cite{Gnedin2001}. 
These reactions occurs in a wide range of matter compositions, 
which includes a strongly-coupled plasma of nuclei and electrons in the outer layers, a 
lattice of neutron-rich nuclei, up to the crust-core interface with abundant free neutrons.
Thus, the resulting emissivities are complicated functions of the temperature and matter composition. 
Considering that the free neutrons in the crust are very likely to pair in the $^1S_0$ state, 
we account for, in addition, the superfluid corrections and the CPFB process in the crust. 
  
Finally, 
we regard relativistic electrons that can
emit neutrino pairs under the presence of a strong magnetic field, which is analogous to 
the synchrotron emission of photons, because our primary goal is to describe the cooling of magnetized NSs. 
This  neutrino synchrotron emissivity  is proportional to the field 
strength \citep{Bez1997} and becomes important  for $B>10^{14}$~G. 

\begin{figure}[hbt]
   \centering
   \includegraphics[angle=-90,width=0.53\textwidth]{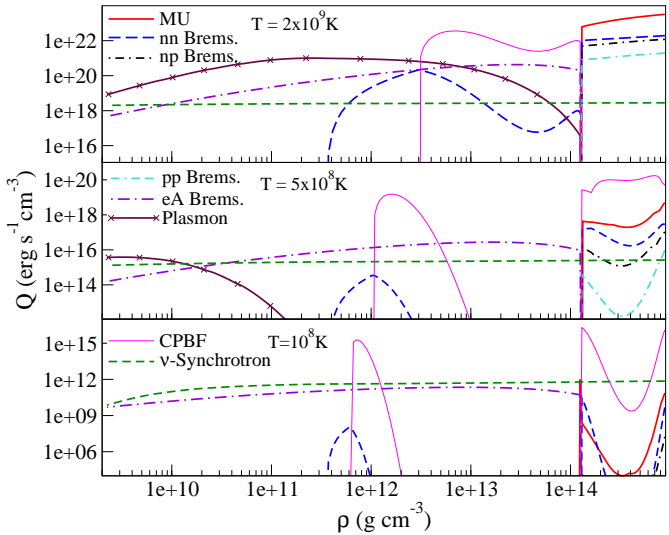}
\caption{Neutrino processes in the crust and in the core for the 
{\em minimal cooling} for different fixed $T$.  Lines denote: MUrca (thick solid), 
$n$-$n$ Bremsstrahlung (long dashed), $n$-$p$ Bremsstrahlung (short dashed dotted), 
$p$-$p$ Bremsstrahlung (double dashed dotted), $e$-$A$ Bremsstrahlung (long dashed dotted), 
Plasmon decay (solid with cross symbols), CPBF (thin solid) and 
$\nu$-Synchrotron (short dashed),  assuming a constant field of $B=10^{14}$ G. 
}
\label{FigNeutrino}
\end{figure}

The emissivities of the most relevant core and crust neutrino processes 
for the {\it minimal cooling} scenario 
are plotted in Fig.~\ref{FigNeutrino},
for three fixed temperatures of $3\times 10^9$~K, $5 \times 10^8$~K, and $1 \times 10^8$~K. 

At $T=3\times 10^9$~K, in the early evolution,  the plasmon decay dominates 
the neutrino emission in the crust and the MUrca  is the strongest energy 
loss mechanism in the core, as seen in the upper panl of Fig.~\ref{FigNeutrino}.  
At this temperature,  neutron superfluidity already exists in the crust, and 
CPBF becomes important near the crust-core interface. 
On the other hand, protons and neutrons in the core have not yet 
condensed into a paired state in a significant volume in the core. 
Later, at intermediate temperatures of 
$T=5\times10^8$~K (middle panel), plasmon decay is dominant 
only in the outer crust, while electron-nuclei Bremsstrahlung becomes more efficient in 
a large part of the crust volume. 
In addition, there is an enhancement of the emissivities due to the CPBF, at densities between 
$10^{13}$-$10^{14}$ g~cm$^{-3}$.
In the core, $^1S_0$ proton superconductivity and $^3P_2$ neutron superfluidity have a 
twofold effect: suppression
of the otherwise dominant processes (MUrca and $N$-$N$ Bremsstrahlung), 
and enhanced emissivity from CPBF.
At later stages, when the temperature has fallen to  $T=10^8$~K, 
neutrino synchrotron 
overcomes the other emissivities if a magnetic field of the order of $B\simeq 10^{14}$ G is present.
A narrow density window is still controlled by CPBF of neutrons in the crust.



\section{Cooling of weakly magnetized neutron stars}
 
Our discussion in based on two baseline models (see Table~\ref{NSmasses})
that correspond to the {\it minimal} and
{\it enhanced} cooling scenarios. The first case 
corresponds to low mass NSs 
in which the central density is below 
the critical density for the onset of the DUrca process, which is   
$2.6 \times 10^{15}$ g~cm$^{-3}$ for our EoS.
The second case 
describes the thermal evolution of a high mass star for 
which the DUrca process operates in a finite volume in the core.
For both models, we solved the thermal diffusion Eq.~(\ref{eneq})
using several magnetic field configurations described in Sect.~\ref{SecMagnetic} and 
the microphysics inputs presented in Sect.~\ref{SecMicrophysics} including the effects of 
superfluidity. We use a two-dimensional numerical grid containing 350 radial and 
60 angular points.

\subsection{Crust formation}

We address the timescale for both the crust formation and the growth of the core 
region where protons are in a superconducting state, because the temperature of a NS
falls below $ 10^{10}$K a few minutes after birth.  
The comparison of these two timescales is relevant to understand whether or not
there is enough time to expel magnetic flux from the core before the crust is formed and the
magnetic field is {\it frozen} into the solid lattice.  If this is the case, 
after the crust is formed, the problem can be treated by assuming
that the magnetic field evolves independently in the crust, without penetrating the core,
while the thermal evolution of the core and the crust become coupled.
In contrast, if a substantial part of the magnetic flux remains
within the core, it is probably organized into superconducting flux tubes that have a
complex interaction with the normal phase. This would be a much more difficult problem to solve and
the evolution would depend on the interaction between the flux tubes and vortices and how they become attached
to the lattice. The study of such a scenario is beyond the scope of this paper, 
and, for simplicity,  we
assume that either the magnetic field has been completely expelled from the core or that 
it penetrates into the core without considering superconducting effects.

\begin{figure}[hbt]
\centering
\includegraphics[angle=-90,width=0.5\textwidth]{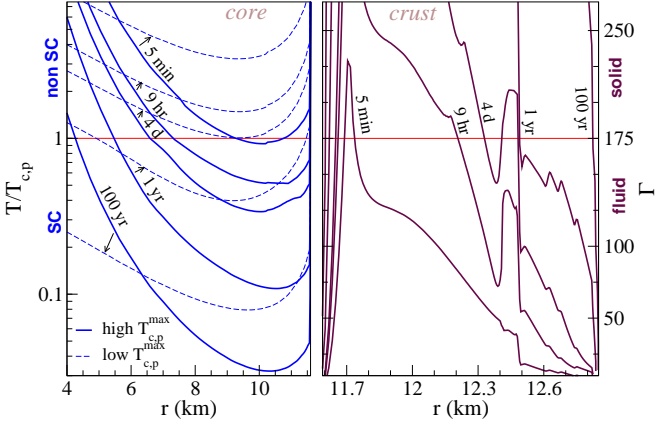}
\caption{Left panel: Growth of the proton superconducting region in the core.
$T/T_{\rm c,p}$ as a function of $r$ at fixed evolution times.  
Solid (dashed) lines correspond to 
high (low) $T_{c,p}^{\rm max}$  for $p$~$^1S_0$, 
i.e. case $e$ ($f$).  
Right panel: Crust formation. $\Gamma$ vs $r$ 
at fixed evolution times. 
In both pannels the LM model (minimal cooling) is used.}.  
\label{FigGamma_tau}
\end{figure}

We followed two indicators of the growth of the crust and the superconducting core:

i) the Coulomb parameter, that describes the physical state of the ions, defined as
\bea
\Gamma = \frac{(Z e)^{2}}{k T a_{i}} \approx
\frac{0.23~Z^{2}}{T_{6}} \left( \frac{\rho}{A} \right)^{1/3}
\label{gamma}
\eea
where $a_{i} = (3/4 \pi n_{i})^{(1/3)}$ is the ion-sphere radius, 
$n_{i}$ is the ion number density, and $T_6$ is the
temperature in units of $10^6$~K.
When $\Gamma < 1$, the ions form a Boltzmann gas, when $1 \leq \Gamma < 175$ their state is
a coupled Coulomb liquid, and when $\Gamma \geq 175$ the liquid freezes into a Coulomb lattice.
The melting temperature ($T_{\rm m}$) for a body-centered cubic (bcc) lattice 
corresponds to the value at which $\Gamma = 175$.
For $\rho=10^{14}$ g~cm$^{-3}$, we have that  $T_{\rm m} \simeq 3 \times 10^{10}$~K , and
the inner crust begins to form  at very early stages of evolution.  
We show the evolution of $\Gamma$ for the LM model
in the right panel of Fig.~\ref{FigGamma_tau}, where each line 
corresponds to a different time.  
The inner crust, up to a radius of $12.4$ km, has formed completely on a timescale of several hours 
to a few days. To form the outer crust, however, takes much longer, about 1-100 yr.    
The solidification depends, in principle, on the matter composition, but we obtained 
similar qualitative results after varying the EoS.

ii) the dimensionless parameter $T/T_{c,p}$ for $^1S_0$ proton pairing;
its evolution describes the growth of the superconducting region in the core,
since for $T \leq T_{c,p}$ protons become superfluid. 
We compare two pairing models taken from Table~\ref{Tablegaps}: 
case $e$ with high $T_{c,p}^{\rm max}$  ($\simeq 7\times 10^{9}$~K)  and 
case $f$ with low $T_{c,p}^{\rm max}$   ($\simeq 2\times 10^9$~K).
The evolution of $T/T_{c,p}$ is shown in the left panel of Fig.~\ref{FigGamma_tau}. 
We found that a large part of the core becomes superconducting on a timescale that 
varies from several days to months, depending on the pairing model.

Consequently, we found similar timescales for the formation of the solid crust and for the 
growth of the
superconducting core. Although it would be interesting to investigate how these two processes 
compete, it is beyond the scope of this work. We assume that our initial 
configuration is a NS with a magnetic field that remains fixed after 
the first few days, which is much shorter than the overall cooling evolution time.  

\begin{figure}[bth]
   \centering
   \includegraphics[angle=-90,width=0.5\textwidth]{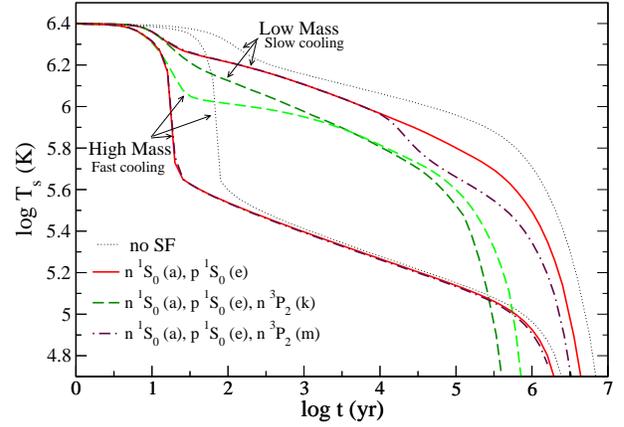}
\caption{Cooling curves of weakly magnetized NS with $B \leq 10^{12}$ G. 
Surface temperature ($T_s$) vs age ($t$) for   LM  and HM stars.
$n$ $^1S_0$ and $p$ $^1S_0$ are fixed to case $a$ and $e$, respectively. 
Solid lines show the case without $n$$^3P_2$ gap; for dashed lines $n$$^3P_2$ 
is fixed to case $k$ 
and for dotted dashed lines to case $m$. Dotted lines represent 
no superfluidity.}
\label{Fig_B0}
\end{figure}
 \subsection{Minimal and Enhanced cooling}

To evaluate whether all microphysics inputs are implemented properly in our
two-dimensional code, we revisit cooling curves for weakly magnetized neutron stars, 
i.e. with field strengths $B \leq 10^{12}$ G.  We compare our results 
 considering that, for weak fields, the temperature profiles are almost spherically symmetric, 
with previous one-dimensional calculations performed by other authors.
The most important deviations between models
arise, as expected, from the underlying microphysics. Major differences depend
on the occurrence of superfluidity and whether  slow or fast neutrino emission processes
are taking place. We summarize these effects below. 

We plot the surface temperature $T_s$ 
for the minimal and enhanced cooling 
scenarios in Fig.~\ref{Fig_B0}.
In both cases, we explored different superfluidity models. The major uncertainties 
come from the $n$~$^3P_2$ pairing gap in the core; furthermore, this gap is expected to have
the strongest impact on the luminosities \citep{Page2004}. 
Hereafter, we  fix two models of the 
$n$~$^1S_0$ superfluidity in the crust and the $p$~$^1S_0$ 
in the core to the cases $a$ and $e$, respectively. 
We checked that replacing them with the other options listed in 
Table~\ref{Tablegaps}, produces slight deviations in the cooling curves. 
We only vary the gap model of the  $n$~$^3P_2$ state between 
three different limit cases: no pairing (no SF), 
case $h$ for high $T_{c,n}^{\rm max}$  ($\simeq 6\times 10^{9}$~K), and 
case $m$ for low $T_{c,n}^{\rm max}$  ($\simeq 2\times 10^{8}$~K). 

In the early stages of evolution (up to $\simeq 10^2$ yr) during the initial 
{\em thermal relaxation of the crust}, the main effect of superfluidity 
is the suppression of  the specific heat of 
free neutrons in the crust (see Fig.~\ref{FigCv}), which leads to a faster temperature 
decrease compared to the 
nonsuperfluid case (dotted lines). The following epoch (up to $\simeq 10^4$-$10^5$~yr) 
is controlled by neutrino emission from the core. The  MUrca and Bremsstrahlung processes 
(or the DUrca process for model B)
are exponentially suppressed, in addition to the
heat capacities of neutrons and protons in the core. 
Nevertheless, core CPFB is important and acts in the opposite direction, increasing 
the emissivities but inside a narrow density window. The overall effect is a faster cooling 
of the LM star. 
The opposite effect is found for the HM star, 
where  a high $T_{c,n}^{\rm max}$ pairing of the  
$n$ $^3P_2$, which covers all the core density region (case $h$), produces a significantly 
higher $T_s$ with respect to the non-superfluid case. If the pairing has a  
low $T_{c,n}^{\rm max}$ (case $m$)
or does not occupy the whole core volume, then the DUrca process is as efficient as 
in the nonsuperfluid case, leading to a very rapid cooling.
In the later cooling phase (from $10^5$-$10^6$ yr), 
when photon luminosity gradually  overtakes the 
neutrino luminosity, the most important effect of superfluidity is the reduction 
of the core specific heat that makes the star cool faster.  

In brief, we confirm all previous results for the cooling of non-magnetized neutron stars and 
do not find any qualitative difference from earlier works
\citep{Yakovlev2004,Page2006}.

\section{Cooling of strongly magnetized neutron stars}

We study now magnetized NSs with $B\geq 10^{13}$~G. 
After analyzing the effect of superfluidity on the cooling curves, we restrict further 
study of magnetized neutron stars to two different limiting scenarios with fixed 
superfluidity models, which can be
summarized as follows:
\begin{enumerate}
\item Model A ({\it minimal cooling}): a LM star with $n$ $^1S_0$ (case $a$) in the crust, 
$p$ $^1S_0$ $p$ (case $e$), and $n$ $^3P_2$ (case $h$) in the core. 
\item Model B ({\it enhanced cooling}): a HM star with $n$ $^1S_0$  (case $a$) in the crust, 
$p$ $^1S_0$ (case $e$), and non-superfluid neutrons in the core
\end{enumerate}

In this section, the magnetic field given provided by the initial model
is kept fixed throughout the entire evolution. The effect of field decay
is separately discussed in the next section.
As outlined before,
one of the most relevant effects of the magnetic field is to 
reduce the electron thermal conductivity 
across magnetic field lines. 
Therefore, heat is essentially forced to flow along magnetic lines, 
which results in anisotropic temperature distributions. Another important effect is that,
as a consequence of the different thermal conductivities in the crust and the core, their thermal
evolution is not always coupled. This effect is also magnified by the presence of strong fields.
After the initial fast transient in which large gradients are allowed due to the reduced
thermal conductivity at high temperature, there are different possible evolutions depending on
the magnetic field geometry.
Since the field lines close to the poles are essentially radial, in general 
the magnetic poles are thermally connected with the core and reflect its temperature.
In contrast, the equator is insulated by a {\it magnetically-induced thermal wall} due to 
large tangential components. Its evolution is thus almost independent of that of the core.
We discuss first our results for purely poloidal fields and then consider the effect
of toroidal fields.
 
\subsection{Purely poloidal magnetic fields}

\begin{figure}[thb]
   \centering
   \includegraphics[angle=-90,width=0.5\textwidth]{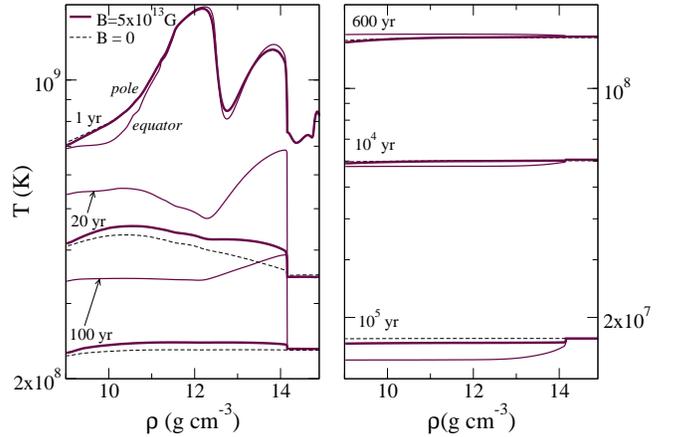}\\
\caption{Evolution of the temperature as a function of the density 
of a LM magnetized NS with $B= 5\times 10^{13}$~G (PC configuration). 
$T$ at the pole is shown with thick solid lines and at the equator with thin solid lines. 
Evolution times are indicated near the lines. The $B=0$ case 
is plotted with dashed lines. 
}
\label{Tprofile}
\end{figure}

In  Fig.~\ref{Tprofile}, we plot temperature profiles across the star, for the
Model A, as a function of the density, and for different evolution times. 
The magnetic field is confined to the crust (PC; $B_r$ and $B_{\theta}$ as in 
Fig.~\ref{Bfield_FF}) and $B=5 \times 10^{13}$~G.
In the early stages, the pole cools down in a similar way to the core and its temperature
is almost identical to that of the non-magnetized  case. 
On the other hand, the equatorial region is decoupled
and shows a different evolution. Since the crustal heat cannot be released inwards,  
 into the core, where neutrino emission is an
efficient cooling mechanism, it remains warmer for a longer time, typically few $10^{2}$ yr 
(Fig.~\ref{Tprofile}, left panel).
At intermediate ages, a nearly isothermal state is reached and the crust and the core evolve 
together, approximately from  $10^{2}$~yr to  $10^{4}$~yr (Fig.~\ref{Tprofile}, right panel). 
At the late evolution ($10^{5}$-$10^{6}$ yr), photon emission from the surface is the 
most efficient way to radiate energy and the initial situation is reversed: since the equator 
cannot be refed by the relatively warmer core, it becomes cooler than the pole. 
We obtained the same qualitative results for Model B. 

\begin{figure}[hbt]
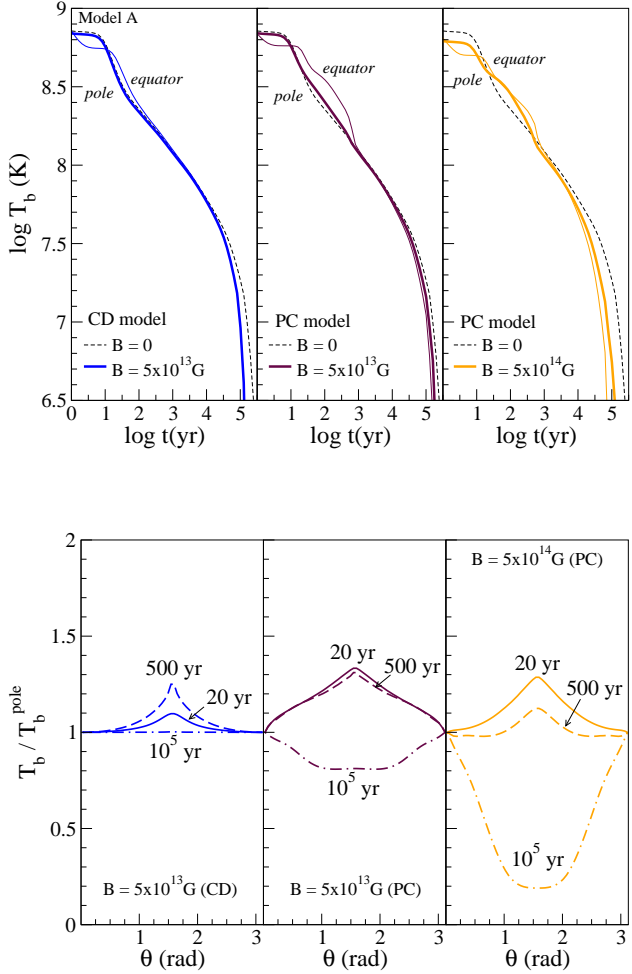

   \centering
\vspace{0.9cm}
 \includegraphics[width=0.45\textwidth]{fig11a.eps}\\
\vspace{1.05cm}
   \includegraphics[width=0.45\textwidth]{fig11b.eps}
\caption{Cooling of strongly magnetized NSs for the Model A. 
Upper panel: $T_b$ vs $t$, at the pole (thick solid lines) and 
at the equator (thin solid lines). Two field configurations are shown: 
CD (left, for $B=5 \times 10^{13}$~G) and PC (middle, 
for $B=5 \times 10^{13}$~G,  and right for $B=5 \times 10^{14}$~G).
Dashed lines indicate the $B=0$ case.
 Lower panel: $T_b/T_b^{\rm pole}$ vs. the azimutal 
angle $\theta$ for three fixed evolution times 
$20$~yr (solid lines), 
$500$~yr  (dashed lines), and
$10^5$~yr (dotted dashed lines).
Similar field configurations as in the upper panel are shown. 
}
\label{BunoA}
\end{figure}

In Fig.~\ref{BunoA}, we show the magnetically induced anisotropic temperature distribution 
for the same model. 
The upper panels are the usual cooling curves (temperature vs. time), which  
display the evolution of $T_b$ at the magnetic pole and the equator, 
for two different field configurations: core dipolar (CD) and poloidal confined 
to the crust (PC). For the latter configuration we show results for two field strengths. 
The cooling curves do not show large deviations between models, although 
the difference with respect to the non-magnetized case becomes larger with
increasing magnetic field strength.
The lower panel
shows the corresponding angular distribution of $T_b$, normalized to its
value at the pole, for three different ages. 
\begin{figure}[hbt]
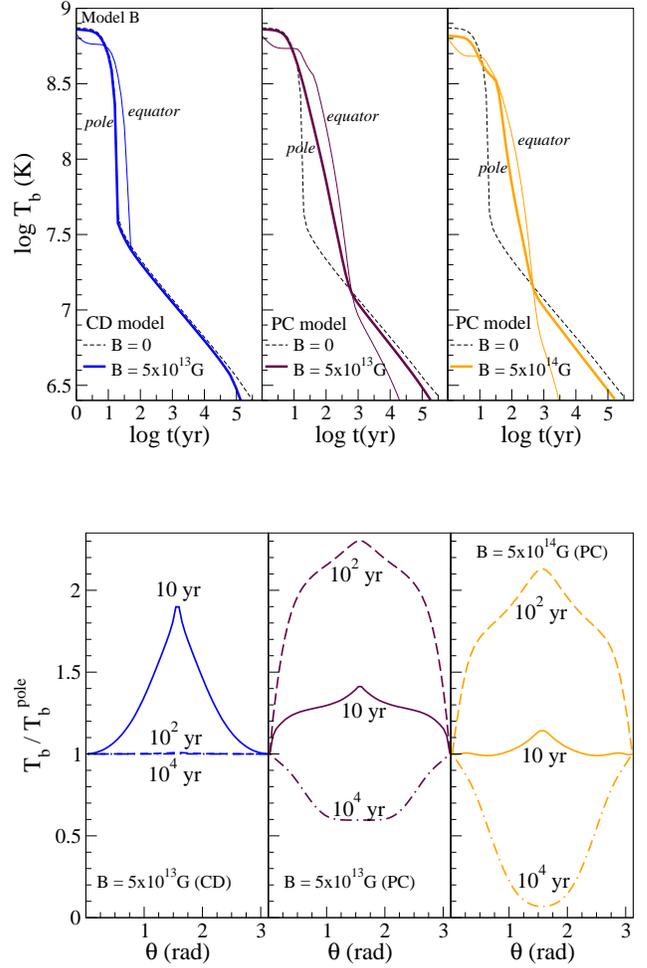

   \centering
\vspace{0.9cm}
   \includegraphics[width=0.45\textwidth]{fig12a.eps}\\
\vspace{1.05cm}
   \includegraphics[width=0.45\textwidth]{fig12b.eps}
\caption{Same as Fig.~\ref{BunoA} but for the Model B. 
}
\label{BunoB}
\end{figure}
At $t \approx 500$ years, we find that, for all models, 
the crustal temperature at the pole is smaller than at the equatorial
region. We refer to an {\em inverted temperature distribution}, in such cases,  when we 
find cooler polar caps with a warmer equatorial belt. 
The occurrence of this {\em inverted} profile is model independent while its duration 
and the degree of anisotropy reached depend on the details of the 
magnetic field geometry and strength.
The equivalent results for Model B are shown in  Fig.~\ref{BunoB}. 
For example, for model B, in which the DU process is allowed, the star cools faster and its 
interior reaches higher values of the magnetization parameter, making the inverted temperature profile 
more pronounced ($T_b^{\rm eq}\simeq 2 T_b^{\rm pole}$). For crustal magnetic fields 
the inverted profile can be maintained for longer times ($\simeq 10^3$ yr) than for the core dipolar 
case that becomes isotropic at about $10^2$ yr (lower panel).  
During late evolution, at $10^5$ yrs, the {\it usual} temperature distribution is found: 
a hot polar cap with a cooler equatorial belt. 

\begin{figure}[hbt]
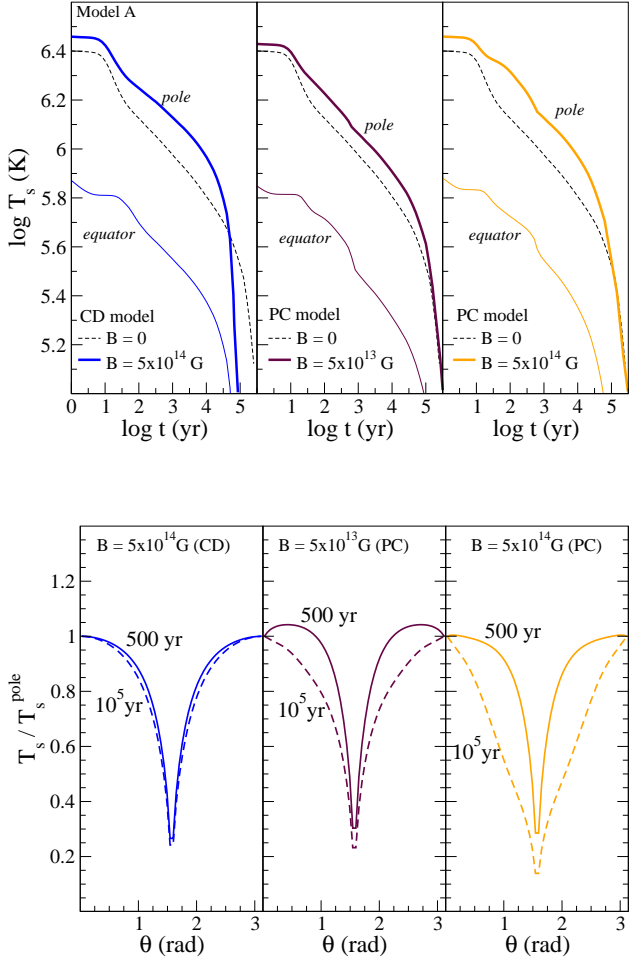

   \centering
\vspace{0.85cm}
\includegraphics[width=0.45\textwidth]{fig13a.eps}\\
\vspace{1.05cm}
\includegraphics[width=0.45\textwidth]{fig13b.eps}
\caption{
Cooling of strongly magnetized NSs for the Model A. 
 Same as Fig.~\ref{BunoA} but for the surface temperature $T_s$ 
(upper panel) and  for
$T_s/T_s^{\rm pole}$ (lower panel).
}
\label{T_surf}
\end{figure}

We reiterate that $T_b$ is the temperature at the bottom of the envelope, corresponding to 
our outer point of integration at $\rho \approx 10^9$ g~cm$^{-3}$, 
and the blanketing effect of the envelope
should be taken into account before comparison with observations. 
To translate the temperature at the base of the envelope to the surface 
temperature $T_s$, we assumed a magnetized envelope as described in Sect.~\ref{BConditions}, 
taking into account the angle that the magnetic field forms with respect to the 
normal to the surface. 
In Fig.~\ref{T_surf}, we plot $T_s$ and $T_s/T^{\rm pole}_s$ vs. age
for the same three cases as in Fig. ~\ref{BunoA}.
We note that the anisotropy found at the level of $T_b$ does not 
automatically produce a similar surface temperature distribution: the blanketing effect of 
the envelope overrides the  {\em inverted temperature distribution} found at
intermediate ages. We see that the equator remains always cooler than the pole, and only
at early times and for strong fields do we find larger surface temperatures
in middle latitude regions. Another important result is that the degree of temperature anisotropy
at the level of the crust is always rather small for magnetic fields penetrating into
the core (i.e. CD), which causes a similar surface temperature distribution
at all times and independently of the magnetic field strength. Crustal confined
fields, in contrast, allow for non-uniform temperatures at the base of the envelope,
leading to temporal variations in the surface temperature distribution during the NS life. 
Since we are interested in the models with the largest variation of temperature,
hereafter we only consider crustal confined fields.

If we compare our temperature angular distribution to former results
\citep{Geppert2004,Azorin2006}, we find that our late time profiles coincide
qualitatively with the stationary solutions obtained in previous works.
However, the temperature distributions at early times are quite different.
The reason is that stationary solutions cannot describe properly the 
temperature distribution in young NSs, because a NS is evolving and changing 
its thermodynamical conditions faster than, or on a similar timescale, to the time 
needed to reach the stationary state.

\subsection{Effect of toroidal fields}

Despite our lack of direct information about the magnetic field geometry inside a neutron star,
there is some agreement that several independent mechanisms can create strong toroidal 
fields, such as differential rotation during core collapse \citep{Wheeler2002}, 
or proto-neutron star dynamo. Hence, it is  natural to investigate the effect
of toroidal components on the surface temperature distribution.
In particular, the inclusion of toroidal components 
was used to explain the small hot emitting areas observed in some isolated NSs
\citep{Perez2006, Geppert2006}. These works concluded that the surface temperature
is determined more by the geometry rather than by the magnetic field strength.
With this motivation, we include a toroidal component in our models and study its influence
on the results.

\begin{figure}[hbt]
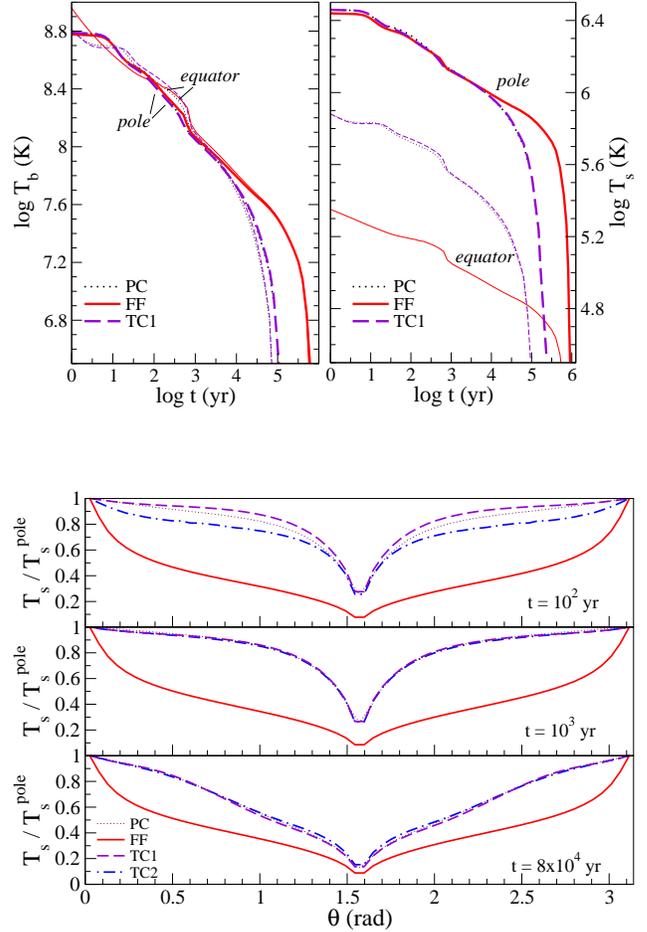

   \centering
\vspace{0.85cm}
\includegraphics[width=0.45\textwidth]{fig14a.eps}\\
\vspace{1.05cm}
\includegraphics[width=0.45\textwidth]{fig14b.eps}
\caption{Cooling of strongly magnetized NSs with toroidal fields. 
Upper panel: $T_b$ vs $t$ (on the left) and $T_s$ vs $t$ 
(on the right) at the pole (thick lines) and 
at the equator (thin lines). 
Three field configurations with $B=5 \times 10^{14}$~G are shown: 
PC (dotted lines), FF (solid lines) and TC1 (dashed lines). 
Lower panel: $T_s/T_s^{\rm pole}$ vs $\theta$ for three fixed evolution times. 
Same field configurations as in the upper panel are shown.  
}
\label{Btor_sup}
\end{figure}

In the remainder of this section, we focus on the effect of the toroidal
component on model A, but our qualitative conclusions can be generally extended to model B. 
In Fig.~\ref{Btor_sup}, we compare the cooling curves in the upper panel, and the
angular temperature distribution in the lower panel, for $B= 5 \times 10^{14}$~G,
for the different toroidal field configurations.
The first conclusion is that the presence of crustal confined toroidal fields (TC1 and TC2)
does not significantly change the results obtained with purely poloidal fields
(dotted lines in Fig.~\ref{Btor_sup}).
We omitted model TC2 in the upper panel because it was indistinguishable 
from TC1; minor differences are visible only during the first 100 years of evolution 
(see lower panel).
We found that the FF configuration exhibits larger $T_b$ than the other
models in the late evolution ($t>10^5$ yr) because the heat transport was 
suppressed by the toroidal component extended through the envelope, and the 
insulating effect was more pronounced. 
The reason for the large differences between models with toroidal magnetic
fields confined to the crust and the FF model, 
was the difference in the angle that the magnetic field forms 
with the normal to the surface. According to Eq.~(\ref{PCY-iron}), the more tangential
the field (the larger $\varphi$), the smaller $T_s$ for a given $T_b$.
In general, we expect that the presence of toroidal fields extended to the
envelope and magnetosphere results in lower surface temperatures in the
equatorial region.  During the neutrino cooling era,
the polar region remains as hot as in the poloidal case, 
but during the photon era, due to the reduced photon luminosity,
the star cools more slowly and the pole remains warmer.

As we can see in the lower panel of Fig.~\ref{Btor_sup} 
for the FF case, the toroidal field maintains, during the entire evolution,
a cooler and more extended
equatorial belt, while the hot polar region is shrunk in comparison to the other models.
Defining the angular size of our {\it polar cap} by the angle at which the radial
component of $B$ becomes larger than the tangential component, this is $B_r^2 > B_{\theta}^2$,
for a purely poloidal configuration,  which implies a hot 
area of about $ \simeq 40^{\circ}-60^{\circ}$. For a FF model, this condition is
reached when $B_r^2 > B_{\theta}^2+B_{\phi}^2$, which  provides a smaller angular size 
of about $ 10^{\circ}$,  which  agrees with the estimated emitting area for 
some isolated neutron stars \citep{Perez2006}. 
The same comment made at the end of the previous subsection about the comparison
with stationary models is valid when comparing these results to stationary temperature
distributions with toroidal fields \citep{Azorin2006,Geppert2006}.

\section{Magnetic field decay}

In the previous section, we discussed the impact of strong magnetic fields
on the cooling and the temperature distribution of NSs, keeping the field strength
and geometry fixed throughout the entire evolution. But the existence of crustal
confined fields supported by crustal currents is inconsistent with the
assumption of non--evolving fields. Currents in the crust dissipate
in a relatively short timescale, which may vary depending on the interaction of
electrons with the lattice in different crustal regions. In any case, 
this leads to dissipation of the magnetic field
on timescales at least comparable, if not shorter, than the cooling timescale.
This effect is important while the crust is still hot because of the large
temperature dependence of the electrical resistivity. At late times, 
that is after a few $10^6$ years, when the crust temperature drops below
$10^7$ K, the conductivity increases significantly, although limited
by electron-impurity or phonon-impurity scattering, and
the magnetic field decays on a much longer timescale.

Since we are interested mostly
in the evolution of NSs while their temperature is sufficiently high for them
to be visible as thermal emitters, the effect of Joule heating by magnetic field 
decay cannot be ignored.

\subsection{Effect of Joule heating on the cooling of magnetized  NSs}
Based on more detailed works studying the magnetic field evolution in NS's crusts
\citep{PonsGeppert2007}, we assumed the simplified form of field decay
provided by Eq.~(\ref{Btime}), and we chose the model TC1 as representative
of the type of fields expected to arise from those simulations.
Although we center our discussion on the particular case of model A with TC1, 
we stress that our results depend qualitatively neither on the particular
NS model (equation of state, crust size, etc.) nor the choice of radial dependence
of the toroidal component.
\begin{figure}[bth]
\vspace{0.85cm}
   \centering
   \includegraphics[width=0.45\textwidth]{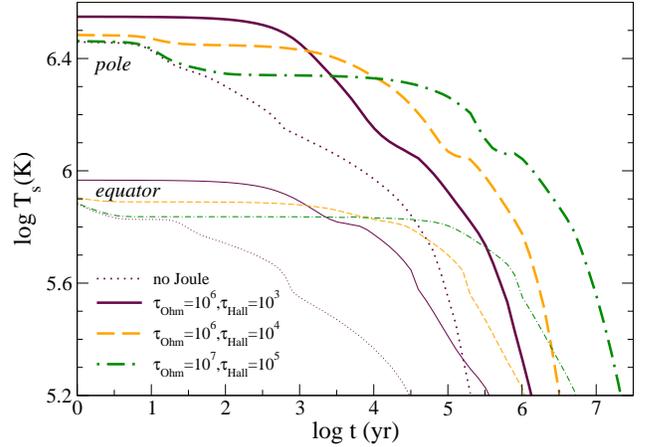}\\
\vspace{1.05cm}
   \includegraphics[width=0.45\textwidth]{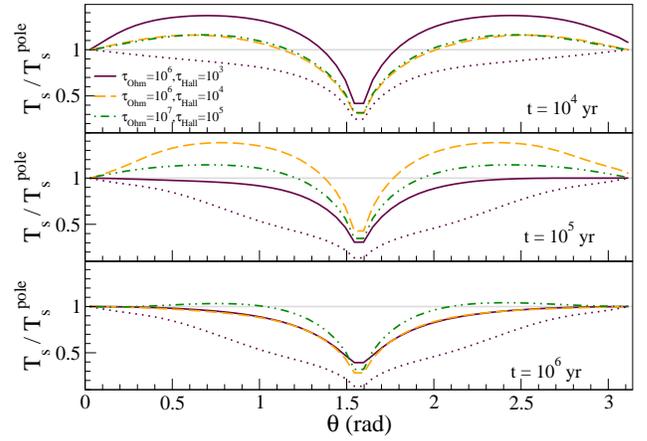}
\caption{Cooling of strongly magnetized NSs with Joule heating with 
$B_0=5 \times 10^{14}~G$. 
Upper panel: $T_s$ vs $t$ at the pole (thick lines) and 
at the equator (thin lines) for 
three pairs 
of values  $(\tau_{\rm Ohm}, \tau_{\rm Hall})$: 
$(10^6, 10^3)$~yr (solid lines), 
$(10^6, 10^4)$~yr  (dashed lines), and
$(10^7, 10^5)$~yr (dotted dashed lines), respectively. 
Lower panel: $T_s/T_s^{\rm pole}$ vs $\theta$ for three fixed evolution times. 
}
\label{JouleA_sup}
\end{figure}

Having fixed our background NS model and field geometry, we varied the parameters
that describe the typical timescales for Ohmic dissipation and a 
fast initial decay induced by the Hall drift.
In Fig.~\ref{JouleA_sup}, we show the cooling curves for three different pairs 
of values  $(\tau_{\rm Ohm}, \tau_{\rm Hall})= \left\{ 
(10^6, 10^3);  (10^6, 10^4);  (10^7, 10^5) \right\}$ yr, represented by
solid lines, dashed lines, and dash-dotted lines, respectively.
For comparison, the dotted lines show the evolution with constant field for
the same initial field ($B_0=5 \times 10^{14}$ G).

The decay of such a large field has an enourmous effect upon the surface temperature;  
due to the heat released, the temperature remains 
far higher than for a non--decaying magnetic field. 
The strong imprint of the field decay is evident for all pairs of parameters chosen. 
We note that the temperature of the initial plateau is higher for shorter $\tau_{\rm Hall}$,
but the duration of this stage, which has almost constant temperature,  is also shorter. 
This is a consequence of releasing a similar amount of heat in a shorter time:  
at $t=\tau_{\rm Hall}$,   $B$ has decayed to about half of its initial value 
and three quarters of the initial magnetic energy has dissipated. By
reducing $\tau_{\rm Hall}$ we can therefore maintain higher temperatures, but for shorter times.
After $t = \tau_{\rm Hall}$, there is a noticeable drop in $T_s$ due to the transition 
from the fast Hall decay to the slower Ohmic decay.

The insulating effect of tangential magnetic fields operates in both directions: in
the absence of additional heating sources, it decouples low latitude regions from the
hotter core resulting in lower temperatures at the base of the envelope; conversely,
if  heat is released in the crust, it prevents extra heat flowing into
the inner crust or the core where it is more easily lost in the form of neutrinos.
Indeed, our simulations that include Joule heating systematically indicate the presence 
of a hot equatorial belt at the crust--envelope interface. 
\cite{Kaminker2006} studied the effect of a localized heat source
at different depths inside a NS. They concluded that
only a heat source very close to the stellar surface can have observational
consequences. In this work, we find evidence for a far more important effect on the surface
temperature. The main reason for this apparent discrepancy is that our cooling
models are two--dimensional and include the insulating effect of the strong
tangential field in the crust, as opposed to the one--dimensional simulations studied by \cite{Kaminker2006}.
However, as discussed in the previous section, this
{\it inverted temperature distribution} at the level of the crust
is not necessarily visible in the surface
temperature distribution because it is filtered by the magnetized envelope.
An analysis of the angular temperature distribution shown in the 
lower panel of Fig.~\ref{JouleA_sup} shows an interesting feature 
related to the heat deposition: the development of a middle latitude region 
hotter than the pole at relatively late stages in the evolution ($t \simeq 10^4, 10^5$ yr).
For a wide range of parameters we found this hot area. It would have implications
for the light curves of rotating NSs, that will differ substantially from the light curves
obtained with a typical model consisting of a hot polar cap with a cooler equatorial region.

\subsection{The hidden direct Urca process ?}

\begin{figure}[bth]
   \centering
   \includegraphics[angle=-90,width=0.5\textwidth]{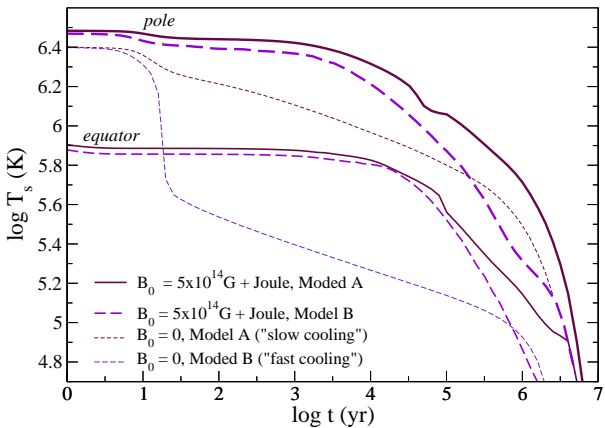}
\caption{Comparison of the fast and slow cooling including Joule heating. 
$T_s$ vs age for Model A (solid lines) and for Model B (dashed lines). The temperature at the pole (thick lines) and at the equator (thin lines) is shown.  
The initial field is $B_0= 5 \times 10^{14} G$ and the field decay rate is of  
$\tau_{\rm Ohm}=10^6$, $\tau_{\rm Hall}=10^4$ yr. 
The $B=0$ case is shown in thin long (short) dashed lines for Model A  (B). 
}
\label{JouleB}
\end{figure}

We conclude our discourse on the important impact of magnetic field decay
in the cooling history of a neutron star by reconsidering the enhanced cooling scenario, 
in which the DUrca efficiently cools the star very quickly.
In Fig.~\ref{JouleB}, we compare our results for the  cooling of low and high mass NS 
(Models  A and B),  with and without magnetic fields. Neglecting the effect of
magnetic fields, the differences between the fast and slow cooling scenarios 
(short dashed and dotted lines, respectively) are clearly
evident, although they can be reduced by strong superfluidity. 
We consider, however, a limiting case that experiments a rapid cooling, which has no superfluidity in the inner core, to observe the significance of the magnetic field effects on the surface temperature (thick solid and dashed lines).
As we can see in Fig.~\ref{JouleB},
in a NS born with a field of $5 \times 10^{14}$~G that decays about one order of magnitude
during the first million years of its life, it is hard to distinguish whether or not a fast neutrino
emission process is active. The surface temperature, both at the pole and the equator,  
is essentially determined by the magnetic field geometry, strength and decay rate.
Only at late times, the differences between models with and without
DUrca become visible, but still the variations between 
models with different field strengths can be larger than the differences stemming from the
fast neutrino cooling process. 
This interesting result could imply that we need to reconsider the observations because a fast neutrino cooling process may well be triggered inside neutron stars but hidden by the magnetic field.
A detailed analysis of the different possibilities of fast cooling (hyperons, quarks,
pure nucleonic matter with large symmetry energy) and comparison with the
observations is beyond the scope of this paper, but the problem is thought-provoking.
Our first results indicate that direct URCA may be veiled
by magnetic fields and this scenario may not be properly identified as
{\em fast cooling} (Aguilera, Pons \& Miralles, 2007, in preparation).

\section{Conclusions}

We have presented a thorough study of the thermal evolution of neutron stars including
some of the most intriguing effects of magnetic fields. Our results were based on two-dimensional
cooling simulations of realistic models that account for the anisotropy in the thermal conductivity
tensor. In the first part of the paper,  we revisited the classical scenario with low magnetic 
fields and presented the input microphysics, working assumptions, and the baseline models. 
As an interesting byproduct, we reconsidered the growth of the crust and of the superconducting
region in the NS core, and found that there are situations in which both growth rates
are comparable. The main body of the work was aimed at the discussion of the two
principal effects of magnetic fields: the anisotropic surface temperature distribution and
the additional heating by magnetic field decay.
We found that, even for purely dipolar fields, 
an inverted temperature distribution is plausible at intermediate ages. Thus the  surface
temperature distribution of neutron stars with high magnetic fields, even in the axisymmetric case,
may be quite different  from the model with two hot polar caps and a cooler equatorial region. The irregular
light curves of some isolated neutron stars, for instance RBS1223, 
\citep{Schwope2005,Kaplan2005} are an indication of such complex structures.  

The main result of this work is that, in NSs born as magnetars, 
Joule heating has an enormous effect on the thermal evolution.  Moreover, this effect is important for
intermediate field stars. If the magnetic field is supported by crustal currents, this effect 
can reach a maximum because two combined factors enhance the efficiency of the heating process:
i) more heat is released into the crust, in the regions of higher resistivity close to the surface, and
ii) large non radial components of the field channel the heat towards the surface,
instead of being lost by neutrinos in the core. As expected,
it becomes clear that magnetic fields and Joule heating are playing a key role keeping
magnetars warm for a long time, but it is likely that the same effect, although quantitatively
smaller, must be considered in radio--quiet isolated NSs or high magnetic field
radio--pulsars.

Another aspect that should be considered when we try to explain observations 
using theoretical cooling curves is  that for many objects the age is estimated
assuming that the loss of angular momentum is entirely due to dipolar radiation
from a magnetic dipole (spin-down age).
In the case of a decaying magnetic field, the {\em spin down age}, 
seriously overestimates the {\em true age} \citep{GO1970}.
Therefore, the cooling evolution time should be corrected,  according
to the prescription for magnetic field decay, 
to compare our model accurately with observations. 
A detailed comparison of the cooling curves  obtained in
this work with observational sources can be found in 
\cite{Aguilera2008}.

Our last striking remark is that the occurrence of direct URCA or, in general, fast neutrino
cooling in NS may be hidden by a combination of effects due to strong magnetic fields.
Our conclusion is that the most appropiate candidates to monitor as rapid coolers are NSs with
fields lower than $10^{13}$ G. Otherwise, we may be misled in our interpretation
of the temperature-age diagrams.

The main drawback of our work is that we are not yet able to return a fully consistent 
simulation of the magneto-thermal coupled evolution of temperature and magnetic field.
In the near future, we plan to extend this study by coupling our thermal diffusion code to
the consistent evolution of the magnetic field in the crust given by the Hall induction 
equation. That approach will permit the accurate evaluation of the heating rates, including 
the non-linear effects associated with the Hall--drift in the NS crust. We believe, however, that
the phenomenological parameterization employed in this paper, reproduces qualitatively the
results expected in a real case.  We have provided another step towards understanding the
cooling of neutron stars, by pointing out a number of important features that must be
more carefully considered in future work.

\begin{acknowledgements}
      This work has been supported by the Spanish MEC grant AYA 2007-67626-C03-C02.  
      D.N.A was supported by VESF-VIRGO project number EGO-DIR-112/2005. 
      We thank U.~Geppert for carefully reading and valuable comments.       
      D.N.A. acknowledges interesting discussions during the INT Neutron Star Crust and Surface 
      Workshop, Seattle. 
\end{acknowledgements}

\bibliography{CMNS_rev1}

\begin{thebibliography}{69}
\expandafter\ifx\csname natexlab\endcsname\relax\def\natexlab#1{#1}\fi

\bibitem[{{Aguilera} {et~al.}(2008){Aguilera}, {Pons}, \&
  {Miralles}}]{Aguilera2008}
{Aguilera}, D.~N., {Pons}, J.~A., \& {Miralles}, J.~A. 2008, \apjl, 673, L167

\bibitem[{{Amundsen} \& {Ostgaard}(1985)}]{Amundsen1985}
{Amundsen}, L. \& {Ostgaard}, E. 1985, Nuclear Physics A, 437, 487

\bibitem[{{Andersson} {et~al.}(2005){Andersson}, {Comer}, \&
  {Glampedakis}}]{Andersson2005}
{Andersson}, N., {Comer}, G.~L., \& {Glampedakis}, K. 2005, Nuclear Physics A,
  763, 212

\bibitem[{{Baiko} {et~al.}(2001){Baiko}, {Haensel}, \& {Yakovlev}}]{Baiko2001}
{Baiko}, D.~A., {Haensel}, P., \& {Yakovlev}, D.~G. 2001, \aap, 374, 151

\bibitem[{{Bailin} \& {Love}(1984)}]{BL84}
{Bailin}, D. \& {Love}, A. 1984, \physrep, 107, 325

\bibitem[{{Baldo} {et~al.}(1998){Baldo}, {Elgar{\o}y}, {Engvik},
  {Hjorth-Jensen}, \& {Schulze}}]{Baldo1998}
{Baldo}, M., {Elgar{\o}y}, {\O}., {Engvik}, L., {Hjorth-Jensen}, M., \&
  {Schulze}, H.-J. 1998, \prc, 58, 1921

\bibitem[{{Bezchastnov} {et~al.}(1997){Bezchastnov}, {Haensel}, {Kaminker}, \&
  {Yakovlev}}]{Bez1997}
{Bezchastnov}, V.~G., {Haensel}, P., {Kaminker}, A.~D., \& {Yakovlev}, D.~G.
  1997, \aap, 328, 409

\bibitem[{{Braithwaite} \& {Spruit}(2004)}]{Braithwaite2004}
{Braithwaite}, J. \& {Spruit}, H.~C. 2004, \nat, 431, 819

\bibitem[{{Canuto} \& {Chiuderi}(1970)}]{Canuto1970}
{Canuto}, V. \& {Chiuderi}, C. 1970, \prd, 1, 2219

\bibitem[{{Chen} {et~al.}(1986){Chen}, {Clark}, {Krotscheck}, \&
  {Smith}}]{Chen1986}
{Chen}, J.~M.~C., {Clark}, J.~W., {Krotscheck}, E., \& {Smith}, R.~A. 1986,
  Nuclear Physics A, 451, 509

\bibitem[{{Chugunov} \& {Haensel}(2007)}]{CH2007}
{Chugunov}, A.~I. \& {Haensel}, P. 2007, \mnras, 381, 1143

\bibitem[{{Cumming} {et~al.}(2006){Cumming}, {Macbeth}, {Zand}, \&
  {Page}}]{Cumming2006}
{Cumming}, A., {Macbeth}, J., {Zand}, J.~J.~M.~i., \& {Page}, D. 2006, \apj,
  646, 429

\bibitem[{{Douchin} \& {Haensel}(2001)}]{Douchin2001}
{Douchin}, F. \& {Haensel}, P. 2001, \aap, 380, 151

\bibitem[{{Elgar{\o}y} {et~al.}(1996{\natexlab{a}}){Elgar{\o}y}, {Engvik},
  {Hjorth-Jensen}, \& {Osnes}}]{Elgaroy1996}
{Elgar{\o}y}, {\O}., {Engvik}, L., {Hjorth-Jensen}, M., \& {Osnes}, E.
  1996{\natexlab{a}}, Physical Review Letters, 77, 1428

\bibitem[{{Elgar{\o}y} {et~al.}(1996{\natexlab{b}}){Elgar{\o}y}, {Engvik},
  {Hjorth-Jensen}, \& {Osnes}}]{Elgaroy1996a}
{Elgar{\o}y}, {\O}., {Engvik}, L., {Hjorth-Jensen}, M., \& {Osnes}, E.
  1996{\natexlab{b}}, Nuclear Physics A, 607, 425

\bibitem[{{Flowers} \& {Itoh}(1976)}]{Flowers1976}
{Flowers}, E. \& {Itoh}, N. 1976, \apj, 206, 218

\bibitem[{{Geppert} {et~al.}(2004){Geppert}, {K{\"u}ker}, \&
  {Page}}]{Geppert2004}
{Geppert}, U., {K{\"u}ker}, M., \& {Page}, D. 2004, \aap, 426, 267

\bibitem[{{Geppert} {et~al.}(2006){Geppert}, {K{\"u}ker}, \&
  {Page}}]{Geppert2006}
{Geppert}, U., {K{\"u}ker}, M., \& {Page}, D. 2006, \aap, 457, 937

\bibitem[{{Geppert} \& {Rheinhardt}(2006)}]{Rheinhardt2006}
{Geppert}, U. \& {Rheinhardt}, M. 2006, \aap, 456, 639

\bibitem[{{Gnedin} \& {Yakovlev}(1995)}]{Gnedin1995}
{Gnedin}, O.~Y. \& {Yakovlev}, D.~G. 1995, Nuclear Physics A, 582, 697

\bibitem[{{Gnedin} {et~al.}(2001){Gnedin}, {Yakovlev}, \&
  {Potekhin}}]{Gnedin2001}
{Gnedin}, O.~Y., {Yakovlev}, D.~G., \& {Potekhin}, A.~Y. 2001, \mnras, 324, 725

\bibitem[{{Goldreich} \& {Reisenegger}(1992)}]{Goldreich1992}
{Goldreich}, P. \& {Reisenegger}, A. 1992, \apj, 395, 250

\bibitem[{{Greenstein} \& {Hartke}(1983)}]{Greenstein1983}
{Greenstein}, G. \& {Hartke}, G.~J. 1983, \apj, 271, 283

\bibitem[{{Gudmundsson} {et~al.}(1983){Gudmundsson}, {Pethick}, \&
  {Epstein}}]{Gudmundsson1983}
{Gudmundsson}, E.~H., {Pethick}, C.~J., \& {Epstein}, R.~I. 1983, \apj, 272,
  286

\bibitem[{{Gunn} \& {Ostriker}(1970)}]{GO1970}
{Gunn}, J.~E. \& {Ostriker}, J.~P. 1970, \apj, 160, 979

\bibitem[{{Haberl}(2007)}]{Haberl2007}
{Haberl}, F. 2007, \apss, 308, 181

\bibitem[{{Haensel} {et~al.}(1996){Haensel}, {Kaminker}, \&
  {Yakovlev}}]{Haensel1996}
{Haensel}, P., {Kaminker}, A.~D., \& {Yakovlev}, D.~G. 1996, \aap, 314, 328

\bibitem[{{Itoh}(1975)}]{Itoh1975}
{Itoh}, N. 1975, \mnras, 173, 1P

\bibitem[{{Itoh} {et~al.}(1984){Itoh}, {Kohyama}, {Matsumoto}, \&
  {Seki}}]{Itoh84}
{Itoh}, N., {Kohyama}, Y., {Matsumoto}, N., \& {Seki}, M. 1984, \apj, 285, 758

\bibitem[{{Jones}(1987)}]{Jones1987}
{Jones}, P.~B. 1987, \mnras, 228, 513

\bibitem[{{Kaminker} {et~al.}(2001){Kaminker}, {Haensel}, \&
  {Yakovlev}}]{Kaminker2001}
{Kaminker}, A.~D., {Haensel}, P., \& {Yakovlev}, D.~G. 2001, \aap, 373, L17

\bibitem[{{Kaminker} {et~al.}(1999){Kaminker}, {Pethick}, {Potekhin},
  {Thorsson}, \& {Yakovlev}}]{Kaminker1999}
{Kaminker}, A.~D., {Pethick}, C.~J., {Potekhin}, A.~Y., {Thorsson}, V., \&
  {Yakovlev}, D.~G. 1999, \aap, 343, 1009

\bibitem[{{Kaminker} \& {Yakovlev}(1994)}]{Kaminker1994}
{Kaminker}, A.~D. \& {Yakovlev}, D.~G. 1994, Astronomy Reports, 38, 809

\bibitem[{{Kaminker} {et~al.}(2006){Kaminker}, {Yakovlev}, {Potekhin},
  {Shibazaki}, {Shternin}, \& {Gnedin}}]{Kaminker2006}
{Kaminker}, A.~D., {Yakovlev}, D.~G., {Potekhin}, A.~Y., {et~al.} 2006, \mnras,
  371, 477

\bibitem[{{Kaplan} \& {van Kerkwijk}(2005)}]{Kaplan2005}
{Kaplan}, D.~L. \& {van Kerkwijk}, M.~H. 2005, \apjl, 635, L65

\bibitem[{{Konenkov} \& {Geppert}(2001)}]{KG2001}
{Konenkov}, D.~Y. \& {Geppert}, U. 2001, Astronomy Letters, 27, 163

\bibitem[{{Lattimer} {et~al.}(1991){Lattimer}, {Prakash}, {Pethick}, \&
  {Haensel}}]{Lattimer1991}
{Lattimer}, J.~M., {Prakash}, M., {Pethick}, C.~J., \& {Haensel}, P. 1991,
  Physical Review Letters, 66, 2701

\bibitem[{{Leinson} \& {P{\'e}rez}(2006)}]{Leinson2006}
{Leinson}, L.~B. \& {P{\'e}rez}, A. 2006, Physics Letters B, 638, 114

\bibitem[{{Levenfish} \& {Yakovlev}(1994)}]{Levenfish1994}
{Levenfish}, K.~P. \& {Yakovlev}, D.~G. 1994, Astronomy Reports, 38, 247

\bibitem[{{Maxwell}(1979)}]{Maxwell1979}
{Maxwell}, O.~V. 1979, \apj, 231, 201

\bibitem[{{Miralles} {et~al.}(1998){Miralles}, {Urpin}, \&
  {Konenkov}}]{Miralles98}
{Miralles}, J.~A., {Urpin}, V., \& {Konenkov}, D. 1998, \apj, 503, 368

\bibitem[{{Misner} {et~al.}(1973){Misner}, {Thorne}, \& {Wheeler}}]{Misner1973}
{Misner}, C.~W., {Thorne}, K.~S., \& {Wheeler}, J.~A. 1973, {Gravitation} (San
  Francisco: W.H.~Freeman and Co., 1973)

\bibitem[{{Page}(1995)}]{Page1995}
{Page}, D. 1995, \apj, 442, 273

\bibitem[{{Page} {et~al.}(2006){Page}, {Geppert}, \& {Weber}}]{Page2006}
{Page}, D., {Geppert}, U., \& {Weber}, F. 2006, Nuclear Physics A, 777, 497

\bibitem[{{Page} {et~al.}(2000){Page}, {Geppert}, \& {Zannias}}]{Page2000}
{Page}, D., {Geppert}, U., \& {Zannias}, T. 2000, \aap, 360, 1052

\bibitem[{{Page} {et~al.}(2004){Page}, {Lattimer}, {Prakash}, \&
  {Steiner}}]{Page2004}
{Page}, D., {Lattimer}, J.~M., {Prakash}, M., \& {Steiner}, A.~W. 2004, \apjs,
  155, 623

\bibitem[{{P{\'e}rez-Azor{\'{\i}}n}
  {et~al.}(2006{\natexlab{a}}){P{\'e}rez-Azor{\'{\i}}n}, {Miralles}, \&
  {Pons}}]{Azorin2006}
{P{\'e}rez-Azor{\'{\i}}n}, J.~F., {Miralles}, J.~A., \& {Pons}, J.~A.
  2006{\natexlab{a}}, \aap, 451, 1009

\bibitem[{{P{\'e}rez-Azor{\'{\i}}n}
  {et~al.}(2006{\natexlab{b}}){P{\'e}rez-Azor{\'{\i}}n}, {Pons}, {Miralles}, \&
  {Miniutti}}]{Perez2006}
{P{\'e}rez-Azor{\'{\i}}n}, J.~F., {Pons}, J.~A., {Miralles}, J.~A., \&
  {Miniutti}, G. 2006{\natexlab{b}}, \aap, 459, 175

\bibitem[{{Pons} \& {Geppert}(2007)}]{PonsGeppert2007}
{Pons}, J.~A. \& {Geppert}, U. 2007, \aap, 470, 303

\bibitem[{{Pons} {et~al.}(2007){Pons}, {Link}, {Miralles}, \&
  {Geppert}}]{PonsLink2007}
{Pons}, J.~A., {Link}, B., {Miralles}, J.~A., \& {Geppert}, U. 2007, Physical
  Review Letters, 98, 071101

\bibitem[{{Pons} {et~al.}(2002){Pons}, {Walter}, {Lattimer}, {Prakash},
  {Neuh{\"a}user}, \& {An}}]{Pons2002}
{Pons}, J.~A., {Walter}, F.~M., {Lattimer}, J.~M., {et~al.} 2002, \apj, 564,
  981

\bibitem[{{Potekhin} {et~al.}(2007){Potekhin}, {Chabrier}, \&
  {Yakovlev}}]{Potekhin2007}
{Potekhin}, A.~Y., {Chabrier}, G., \& {Yakovlev}, D.~G. 2007, \apss, 308, 353

\bibitem[{{Potekhin} \& {Yakovlev}(2001)}]{Potekhin2001}
{Potekhin}, A.~Y. \& {Yakovlev}, D.~G. 2001, \aap, 374, 213

\bibitem[{{Raedler}(2000)}]{Raedler2000}
{Raedler}, K.-H. 2000, in Lecture Notes in Physics, Berlin Springer Verlag,
  Vol. 556, From the Sun to the Great Attractor, ed. D.~{Page} \& J.~G.
  {Hirsch}, 101--+

\bibitem[{{Schulze} {et~al.}(1998){Schulze}, {Baldo}, {Lombardo}, {Cugnon}, \&
  {Lejeune}}]{Schulze1998}
{Schulze}, H.-J., {Baldo}, M., {Lombardo}, U., {Cugnon}, J., \& {Lejeune}, A.
  1998, \prc, 57, 704

\bibitem[{{Schwope} {et~al.}(2005){Schwope}, {Hambaryan}, {Haberl}, \&
  {Motch}}]{Schwope2005}
{Schwope}, A.~D., {Hambaryan}, V., {Haberl}, F., \& {Motch}, C. 2005, \aap,
  441, 597

\bibitem[{{Schwope} {et~al.}(2007){Schwope}, {Hambaryan}, {Haberl}, \&
  {Motch}}]{Schwope2007}
{Schwope}, A.~D., {Hambaryan}, V., {Haberl}, F., \& {Motch}, C. 2007, \apss,
  308, 619

\bibitem[{{Shibanov} \& {Yakovlev}(1996)}]{ShibYak1996}
{Shibanov}, Y.~A. \& {Yakovlev}, D.~G. 1996, \aap, 309, 171

\bibitem[{{Thompson} \& {Duncan}(1993)}]{TD1993}
{Thompson}, C. \& {Duncan}, R.~C. 1993, \apj, 408, 194

\bibitem[{{Tr{\"u}mper} {et~al.}(2004){Tr{\"u}mper}, {Burwitz}, {Haberl}, \&
  {Zavlin}}]{Truemper2004}
{Tr{\"u}mper}, J.~E., {Burwitz}, V., {Haberl}, F., \& {Zavlin}, V.~E. 2004,
  Nuclear Physics B Proceedings Supplements, 132, 560

\bibitem[{{Urpin} \& {Yakovlev}(1980)}]{Urpin1980}
{Urpin}, V.~A. \& {Yakovlev}, D.~G. 1980, Soviet Astronomy, 24, 425

\bibitem[{{van Riper}(1991)}]{VanRiper1991}
{van Riper}, K.~A. 1991, \apjs, 75, 449

\bibitem[{{Wambach} {et~al.}(1993){Wambach}, {Ainsworth}, \&
  {Pines}}]{Wambach1993}
{Wambach}, J., {Ainsworth}, T.~L., \& {Pines}, D. 1993, Nuclear Physics A, 555,
  128

\bibitem[{{Wheeler} {et~al.}(2002){Wheeler}, {Meier}, \&
  {Wilson}}]{Wheeler2002}
{Wheeler}, J.~C., {Meier}, D.~L., \& {Wilson}, J.~R. 2002, \apj, 568, 807

\bibitem[{{Yakovlev} {et~al.}(2001){Yakovlev}, {Kaminker}, {Gnedin}, \&
  {Haensel}}]{YakovReport2001}
{Yakovlev}, D.~G., {Kaminker}, A.~D., {Gnedin}, O.~Y., \& {Haensel}, P. 2001,
  \physrep, 354, 1

\bibitem[{{Yakovlev} {et~al.}(1999){Yakovlev}, {Kaminker}, \&
  {Levenfish}}]{Yakovlev1999A}
{Yakovlev}, D.~G., {Kaminker}, A.~D., \& {Levenfish}, K.~P. 1999, \aap, 343,
  650

\bibitem[{{Yakovlev} \& {Levenfish}(1995)}]{YakovlevLevenfish1995}
{Yakovlev}, D.~G. \& {Levenfish}, K.~P. 1995, \aap, 297, 717

\bibitem[{{Yakovlev} \& {Pethick}(2004)}]{Yakovlev2004}
{Yakovlev}, D.~G. \& {Pethick}, C.~J. 2004, \araa, 42, 169

\bibitem[{{Zavlin}(2007)}]{Zavlin2007}
{Zavlin}, V.~E. 2007, ArXiv Astrophysics e-prints, astro-ph/0702426

\end{thebibliography}
\end{document}